\pdfoutput=1
\documentclass[usenatbib,twocolumn]{mnras}

\usepackage[T1]{fontenc}
\usepackage{ae,aecompl}

\usepackage[dvipdfmx]{graphicx}	
\usepackage{amsmath}	
\usepackage{amssymb}	
\usepackage{color}
\usepackage{dcolumn}
\usepackage{bm}
\usepackage{comment}

\newcommand{\beq}{\begin{equation}}
\newcommand{\beqa}{\begin{eqnarray}}
\newcommand{\eeq}{\end{equation}}
\newcommand{\eeqa}{\end{eqnarray}}

\newcommand{\simgt}{\lower.5ex\hbox{$\; \buildrel > \over \sim \;$}}
\newcommand{\simlt}{\lower.5ex\hbox{$\; \buildrel < \over \sim \;$}}

\newcommand{\bd}[1]{\mbox{\boldmath $#1$}}

\RequirePackage{lineno}

\def\apj{{ApJ}}                 
\def\apjl{{ApJ}}                
\def\apjs{{ApJS}}               
\def\ao{{Appl.~Opt.}}           
\def\aap{{A\&A}}                

\def\mnras{MNRAS}             
\def\prd{{Phys.~Rev.~D}}        
\def\physrep{{Phys.~Rep.}}   

\title[Imprint of $f(R)$ gravity on weak lensing I\hspace{-.1em}I ]{
The imprint of $f(R)$ gravity on weak gravitational lensing 
I\hspace{-.1em}I : Information content in cosmic shear statistics
}

\author[M.Shirasaki et al.]
{Masato Shirasaki$^{1}$
\thanks{E-mail:masato.shirasaki@nao.ac.jp},
Takahiro Nishimichi$^{2,3}$,
Baojiu Li$^{4}$,
and Yuichi Higuchi$^{5}$
\\
$^{1}$Division of Theoretical Astronomy, National Astronomical Observatory of Japan, 2-21-1 Osawa, Mitaka, Tokyo 181-8588, Japan\\
$^{2}$Kavli Institute for the Physics and Mathematics of the 
Universe (WPI), The University of Tokyo Institutes for Advanced Study, \\
The University of Tokyo, 5-1-5 Kashiwanoha, Kashiwa 277-85
83, Japan\\
$^{3}$CREST, JST, 4-1-8 Honcho, Kawaguchi, Saitama, 332-0012, 
Japan \\
$^{4}$Institute for Computational Cosmology, Department of Physics, Durham University, South Road, Durham DH1 3LE, UK
\\
$^{5}$Academia Sinica Institute of Astronomy and Astrophysics (ASIAA), No. 1, Sec.4, Roosevelt Rd, Taipei 10617, Taiwan\\
}

\date{Accepted XXX. Received YYY; in original form 3/3/2016}

\pubyear{2016}

\begin{document}
\label{firstpage}
\pagerange{\pageref{firstpage}--\pageref{lastpage}}
\maketitle

\begin{abstract}
We investigate the information content of
various cosmic shear statistics on the theory of gravity.
Focusing on the Hu-Sawicki-type $f(R)$ model, 
we perform a set of ray-tracing simulations and
measure the convergence bispectrum, peak counts and Minkowski functionals.
We first show that while the convergence power spectrum does have sensitivity 
to the current value of extra scalar degree of freedom $|f_{\rm R0}|$,
it is largely compensated by a change 
in the present density amplitude parameter $\sigma_{8}$ and the matter density parameter $\Omega_{\rm m0}$.
With accurate covariance matrices obtained from 1000 lensing simulations, 
we then examine the constraining power
of the three additional statistics.
We find that these probes are indeed helpful to break the parameter degeneracy, 
which can not be resolved from the power spectrum alone.
We show that especially the peak counts and Minkowski functionals 
have the potential to rigorously (marginally) detect the signature of modified gravity 
with the parameter $|f_{\rm R0}|$ as small as $10^{-5}$ ($10^{-6}$) 
if we can properly model them on small ($\sim 1\, \mathrm{arcmin}$) scale 
in a future survey with a sky coverage of 1,500 squared degrees.
We also show that the signal level is 
similar among the additional three statistics and all of them 
provide complementary information to the power spectrum.
These findings indicate the importance of combining multiple probes
beyond the standard power spectrum analysis to detect possible modifications to General Relativity.
\end{abstract}
\begin{keywords}
gravitational lensing: weak, large-scale structure of Universe
\end{keywords}


\section{INTRODUCTION}

General Relativity (GR) is the standard theory of gravity
and plays an essential role for 
astronomy, astrophysics and cosmology.
The theory can provide a reasonable explanation for
various phenomena, e.g., 
the anomalous perihelion precession of Mercury's orbit, 
the deflection of radiation from a distant source 
known as gravitational lensing
\citep[e.g.,][]{1920RSPTA.220..291D, 2009ApJ...699.1395F}, 
the time delay by the time dilation in the gravitational lensing in the Sun
\citep[e.g.,][]{1971PhRvL..26.1132S, 2003Natur.425..374B},
the redshift of light moving in a gravitational field,
\citep[e.g.,][]{1980PhRvL..45.2081V},
the orbital decay of binary pulsars,
\citep[e.g.,][]{1982ApJ...253..908T},
and the propagation of ripples in the curvature of space-time 
measured by the Advanced LIGO detectors
\citep{2016PhRvL.116f1102A}.
Assuming that GR
is the correct theory of gravity even on cosmological scales, 
an array of large astronomical observations 
\citep[e.g.,][]{1997ApJ...483..565P, 2006PhRvD..74l3507T, 2015arXiv150201589P} has established 
the standard cosmological model called the $\Lambda$ cold dark matter ($\Lambda$CDM) model.
Although the $\Lambda$CDM model can provide a remarkable fit 
to various observational results,
the correctness of GR on cosmological scales 
is poorly examined so far.
A simple extension of the $\Lambda$CDM model 
can be realized by modification of GR.
This class of cosmological models is known as modified gravity
which can explain the cosmic acceleration at redshift of $z\simlt1$
without introducing the cosmological constant $\Lambda$.
In order to probe the modification of gravity on cosmological scales,
the measurement of the gravitational growth 
of cosmic matter density would be essential 
because the modification could lead to 
some distinct features from the $\Lambda$CDM model
in the matter distribution in the Universe
\citep[for a review, see e.g.,][]{2012PhR...513....1C}.

$f(R)$ gravity is a type of modified gravity theory
which generalizes GR by introducing
an arbitrary function of the Ricci scalar $R$ 
in the Einstein-Hibert action.
This extension can explain the accelerated expansion,
and the resulting extra scalar degree of freedom can
increase the strength of gravity and enhance structure formation.
The deviation from standard gravity must be suppressed
locally to pass stringent tests of GR in the Solar
system, and this can be achieved by virtue of the
chameleon screening.
Interestingly, viable models of $f(R)$ gravity
predict that gravitational lensing effect is governed by
the same equation as in GR
\citep[e.g.,][]{2010LRR....13....3D}.
Observationally,
gravitational lensing is known as 
a robust probe of the underlying matter distribution 
in the Universe independent of the galaxy-biasing uncertainty.
Thus, such measurements in upcoming imaging surveys
could be a powerful tool to constrain cosmological scenarios governed by $f(R)$ gravity.
Cosmic shear is the small distortion of images of distant sources
originating from the bending of light rays passing through 
the large-scale structure in the Universe.
In practice, image distortion induced by gravitational lensing is
smaller than the intrinsic ellipticity of sources.
Therefore, one needs to analyse the data \textit{statistically}
in order to extract purely cosmological information 
arising from gravitational lensing.
Furthermore, the statistics of the cosmic shear field
significantly deviates from Gaussian, reflecting the non-linearity 
of the structure growth.
This fact means that one can not extract the full information
in cosmic shear by using two-point statistics alone.
Ongoing and future galaxy imaging surveys are aimed 
at measuring the cosmic shear signal with 
a high accuracy over several thousand squared degrees.
Thus, it is important and timely to investigate the information about
$f(R)$ gravity in various cosmic shear statistics for the
purpose of making the best use of galaxy imaging surveys.

In this paper, we perform ray-tracing simulations of 
gravitational lensing in the framework of $f(R)$ gravity 
and explore the cosmological information content in 
four different statistics;
the convergence power spectrum, bispectrum, the abundance of peaks
and the Minkowski functionals (MFs).
The first statistic is the basic quantity in modern cosmology 
and describes the correlation of cosmic shear at two different
directions.
The other three quantities would contain information
that supplement the power spectrum.
They extract \textit{non-Gaussian} aspects of the cosmic shear field
through the correlation at three points, the abundance of massive objects
associated with rare peaks near the edge of the (one-point) distribution
and the morphology of the field, respectively.
These statistics 
have already been measured 
in existing weak lensing surveys \citep[e.g.,][]{2013MNRAS.430.2200K, 2014MNRAS.441.2725F, 2014ApJ...786...43S, 2015PhRvD..91f3507L} 
and their usefulness in cosmological analyses have also been 
demonstrated theoretically~\citep[e.g.,][]{2003MNRAS.340..580T, 2004MNRAS.350..893H, Kratochvil2012, 
2012ApJ...760...45S, 2012A&A...541A.161V}. 
We extend the previous analyses of cosmic shear to modified gravity scenarios governed by $f(R)$ gravity
using numerical simulations and testing their statistical power to constrain the parameter in the model.

This paper is organized as follows. 
In Section~\ref{sec:model}, we briefly describe the cosmological model 
based on $f(R)$ gravity and the characteristics of the model.
In Section~\ref{sec:WL}, we summarize the basics of weak lensing 
and cosmic shear statistics used in this paper.
We also explain the details of our lensing simulation 
and the methodology to measure cosmic shear statistics 
in Section~\ref{sec:sim_and_ana}.
In Section~\ref{sec:res}, we provide results of our lensing analysis 
in numerical simulation of modified gravity and compare the results 
between the $f(R)$ model and the $\Lambda$CDM model in detail.
We then quantify the information on the deviation from GR in 
cosmic shear statistics and compare among different statistics.
Conclusions and discussions are presented in Section~\ref{sec:con}.

\section{COSMOLOGICAL MODEL}\label{sec:model}

In this paper, we study a class of cosmological models 
with modified gravity called $f(R)$ gravity.
This model can explain the observed cosmic acceleration at $z\simlt1$
without introducing the cosmological constant
and satisfy the Solar system tests with appropriate parameters.

\subsubsection*{f(R) model}

In $f(R)$ model, 
a general function of the scalar curvature $R$ 
is introduced in the Einstein-Hilbert action
\citep{2006hep.th....1213N,  2010LRR....13....3D, 2015MNRAS.452.3179S};
\beqa
S_{G} = \int {\rm d}^{4}x \sqrt{-g}\left[\frac{R+f(R)}{16\pi G}\right], \label{eq:action_fR}
\eeqa
where $g$ is the determinant of metric and 
$G$ represents the gravitational constant.
The action in Eq.~(\ref{eq:action_fR}) leads to the modified Einstein equation as
\beqa
G_{\mu \nu} + f_{R}R_{\mu \nu}-\left(\frac{f}{2} - \Box f_{R} \right)g_{\mu \nu}
-\nabla_{\mu}\nabla_{\nu}f_{R} = 8\pi G T_{\mu \nu},
\eeqa
where $f_{R} \equiv {\rm d}f/{\rm d}R$, $G_{\mu \nu}\equiv R_{\mu \nu}-1/2g_{\mu \nu}R$ and $\Box\equiv\nabla^\alpha\nabla_\alpha$.
Assuming a Friedmann-Robertson-Walker (FRW) metric, 
one can determine the time evolution of the Hubble parameter in 
$f(R)$ model as follows:
\beqa
H^{2}-f_{R}\left(H\frac{{\rm d}H}{{\rm d}\ln a}+H^2 \right)
+\frac{f}{6}+H^{2}f_{RR}\frac{{\rm d}R}{{\rm d}\ln a} = \frac{8\pi G}{3}\bar{\rho}_{m}, \label{eq:H_fR}
\eeqa
where $a$ is the scale factor and $H = a^{-1}{\rm d}a/{\rm d}t$.
Structure formation in $f(R)$ gravity is governed by 
the modified Poisson equation and the equation of motion 
for the additional scalar degree of freedom $f_{R}$\footnote{
Throughout this paper, we work with the quasi-static approximation. 
\citet{2008PhRvD..77l3515D, 2015JCAP...02..034B} have shown that 
the quasi-static approximation becomes quite reasonable
for models with $|f_{R}| \ll 1$ today. 
}:
\beqa
\nabla^2 \Phi &=&
\frac{16\pi G}{3}\delta \rho_{m}a^2 - \frac{a^2}{6}\delta R, 
\label{eq:poisson_eq_fR} \\
\nabla^2 \delta f_{R} &=&
\frac{a^2}{3}\left[ \delta R - 8\pi G \delta \rho_{m}\right],
\label{eq:field_eq_fR}
\eeqa
where $\Phi$ is the gravitational potential, 
$\delta f_{R} = f_{R}(R)-f_{R}(\bar{R})$,
$\delta R = R - \bar{R}$,
$\delta \rho_{m} = \rho_{m}-\bar{\rho}_{m}$,
and we represent the background quantity with a bar.
Eqs.~(\ref{eq:poisson_eq_fR}) and (\ref{eq:field_eq_fR})
show two notable features in $f(R)$ gravity.
In the high curvature limit where $R \rightarrow 8\pi G \rho_{m}$,
the extra scalar degree of freedom $f_{R}$ 
in Eq.~(\ref{eq:field_eq_fR}) would vanish
and Eq.~(\ref{eq:poisson_eq_fR}) can reproduce the Poisson equation in GR as $\nabla^2 \Phi = 4\pi G a^2 \delta \rho_{m}$.
This is known as the chameleon mechanism required to recover 
GR in high density region \citep[e.g.,][]{2004PhRvD..69d4026K}.
On the other hand, $f_{R}$ would operate in the low curvature regime
where $R < 8\pi G \rho_{m}$ and 
Eq.~(\ref{eq:poisson_eq_fR}) can be approximated by 
$\nabla^2 \Phi = 16 \pi G/3 a^2 \delta \rho_{m}$ 
in the limit of $R \ll 8\pi G \rho_{m}$, 
making the gravity enhanced by a factor of $1/3$.
Therefore, the gravitational force in $f(R)$ model can be enhanced 
depending on the local density environment.

In this paper, we will consider the representative example 
of $f(R)$ models as proposed in \citet[][hereafter denoted as HS model]{2007PhRvD..76f4004H},
\beqa
f(R) = -2\Lambda \frac{R^{n}}{R^{n}+\mu^{2n}},
\eeqa
where $\Lambda$, $\mu$ and $n$ are free parameters in this model.
The sign of $f(R)$ is determined by the condition ${\rm d}^2 f/{\rm d}R^2 >0$
to ensure that the evolution of linear perturbations is 
stable at high-curvature (i.e., no tachyonic instability; 
\citet{2007PhRvD..75d4004S}).
Although the model does not contain a cosmological constant 
as $R\rightarrow0$ (or the limit of flat space-time),
one can approximate the function of $f(R)$ as follows for $R\gg \mu^2$:
\beqa
f(R) = -2\Lambda -\frac{f_{R0}}{n}\frac{\bar{R}_{0}^{n+1}}{R^n},
\eeqa
where $\bar{R}_{0}$ is the present scalar curvature of 
the background space-time and 
$f_{R0} = -2\Lambda \mu^2/\bar{R}_{0}^2 = f_{R}(\bar{R}_{0})$.
In the following, we focus on the case of $n=1$.
In the HS model with $|f_{R0}| \ll 1$, 
the background expansion is almost 
equivalent to that in the $\Lambda$CDM model.
Therefore, in practice, 
geometric tests such as distance measurement with supernovae 
could not distinguish between the $\Lambda$CDM model
and the HS model for $|f_{R0}| \ll 10^{-2}$
\citep{2012PhRvD..85b4006M}.
It is thus of great importance to have other probes to break this degeneracy at the background level.
A natural choice for this is the measurement of gravitational structure growth.
Indeed, Eqs.~(\ref{eq:poisson_eq_fR}) and (\ref{eq:field_eq_fR}) indicate that the signature of modified gravity might exist in the evolution of perturbations.

The evolution of density perturbations in the HS model has been investigated with analytic \citep[e.g.,][]{2007PhRvD..75f4020B} 
and numerical approaches \citep[e.g.,][]{2008PhRvD..78l3524O, 2009PhRvD..79h3518S, 2011PhRvD..83d4007Z, 2013MNRAS.428..743L, 2013PhRvD..88j3507H, 2014ApJS..211...23Z}.
The matter density perturbations in the linear regime 
is scale-dependent as opposed to GR, 
while the nonlinear gravitational growth can be even
more complicated than that in the $\Lambda$CDM model 
(e.g., the chameleon mechanism operates in high-density regions and the $\Lambda$CDM-like gravity should be recovered in such regions). 
Hence, a detailed investigation of matter density distribution in the Universe would be useful to constrain modification of gravity 
due to $f_{R}$.
Note that cosmic shear is among the interesting observables to measure 
matter density distribution in an unbiased way.
 
\section{WEAK LENSING}\label{sec:WL}

We first summarize the basics of gravitational lensing 
induced by large-scale structure.
Weak gravitational lensing effect is usually characterized by
the distortion of image of a source object by the 
following 2D matrix:
\beqa
A_{ij} = \frac{\partial \beta^{i}}{\partial \theta^{j}}
           \equiv \left(
\begin{array}{cc}
1-\kappa -\gamma_{1} & -\gamma_{2}-\omega  \\
-\gamma_{2}+\omega & 1-\kappa+\gamma_{1} \\
\end{array}
\right), \label{distortion_tensor}
\eeqa
where we denote the observed position of a source object as $\bd{\theta}$ 
and the true position as $\bd{\beta}$,
$\kappa$ is the convergence, $\gamma$ is the shear,
and $\omega$ is the rotation.
In the weak lensing regime (i.e., $\kappa, \gamma \ll 1$), 
each component of $A_{ij}$ can be related to
the second derivative of the gravitational potential $\Phi$ as
\beqa
A_{ij} &=& \delta_{ij} - \Phi_{ij}, \label{eq:Aij} \\
\Phi_{ij}  &=&\frac{2}{c^2}\int _{0}^{\chi}{\rm d}\chi^{\prime} f(\chi,\chi^{\prime}) 
\frac{\partial^2}{\partial x_{i}\partial x_{j}}
\Phi[r(\chi^{\prime})\bd{\theta},\chi^{\prime}], \label{eq:shear_ten}\\
f(\chi,\chi^{\prime}) &=& \frac{r(\chi-\chi^{\prime})r(\chi^{\prime})}{r(\chi)},
\eeqa
where 
$\chi$ is the comoving distance,
$r(\chi)$ is the angular diameter distance, 
and $x_{i}=r\theta_{i}$ represents the physical distance 
\citep[]{2001PhR...340..291B}.
By using the Poisson equation and the Born approximation
\citep[]{2001PhR...340..291B}, 
one can express the weak lensing convergence field as
\beqa
\kappa(\bd{\theta},\chi)= \frac{3}{2}\left(\frac{H_{0}}{c}\right)^2 \Omega_{\rm m0}
\int _{0}^{\chi}{\rm d}\chi^{\prime} f(\chi,\chi^{\prime}) 
\frac{\delta[r(\chi^{\prime})\bd{\theta},\chi^{\prime}]}{a(\chi^{\prime})}. \label{eq:kappa_delta}
\eeqa
In general, the lensing equation is governed by the so-called lensing potential $(\Phi+\Psi)/2$ where 
$\Phi$ and $\Psi$ are the Bardeen potentials appearing in the metric perturbation in the Newtonian gauge.
The lensing potential in $f(R)$ gravity would be governed by 
the same Poisson equation as in GR, making 
Eqs.~(\ref{eq:Aij}), (\ref{eq:shear_ten}) and 
(\ref{eq:kappa_delta}) applicable in the HS model with $|f_{R0}| \ll 1$
(the derivation can be found in e.g., \citealt{2014MNRAS.440..833A}).
In this paper, we take into account 
the non-linearity of the convergence field 
entering in Eq.~(\ref{eq:shear_ten}) 
using the ray-tracing technique over simulated density fields.

\subsection{Cosmic shear statistics}
We here introduce four different 
statistics of the cosmic shear.
In this paper, we consider statistical analysis with 
the convergence power spectrum, bispectrum,
peak counts and MFs.
The power spectrum has complete cosmological information 
when the fluctuation follows the Gaussian statistics.
However, the nonlinear structure formation induced by gravity 
induces non-Gaussianity even if the initial fluctuations are Gaussian distributed.
Therefore, higher-order statistics can be important to fully exploit 
weak lensing maps beyond the power spectrum analysis.

\subsubsection{Power spectrum}
The power spectrum is one of the basic statistics in modern cosmology 
\citep[e.g.,][]{2012MNRAS.427.3435A, 2015arXiv150702704P, Becker:2015ilr}.
It is defined as the two-point correlation in Fourier space.
In the case of the convergence field $\kappa$, that is
\begin{eqnarray}
\langle \tilde{\kappa}(\bm{\ell}_1) \tilde{\kappa}(\bm{\ell}_2) \rangle 
= (2\pi)^2 \delta_D (\bm{\ell}_{1}+\bm{\ell}_{2})P_\kappa (\ell_1),
\end{eqnarray}
where $\delta_{D}(\bm{x})$ is the Dirac delta function and 
the multipole $\ell$ is related to the angular scale through $\theta=\pi/\ell$. 
By using the Limber approximation \citep{Limber:1954zz,Kaiser:1991qi}
and Eq.~\eqref{eq:kappa_delta}, 
one can express the convergence power spectrum as 
\beqa
P_{\kappa}(\ell) = \int_{0}^{\chi_s} {\rm d}\chi \frac{W(\chi)^2}{r(\chi)^2} 
P_{\delta}\left(k=\frac{\ell}{r(\chi)},z(\chi)\right)
\label{eq:kappa_power},
\eeqa
where $P_{\delta}(k)$ represents the three dimensional matter power spectrum, 
$\chi_s$ is the comoving distance to the source galaxies 
and $W(\chi)$ is the lensing weight function defined as
\beqa
W(\chi) = \frac{3}{2}\left(\frac{H_{0}}{c}\right)^2 
\Omega_{\rm m0}
\frac{r(\chi_s-\chi)r(\chi)}{r(\chi_s)}(1+z(\chi)),
\eeqa
where 
$H_{0}$ is the present-day Hubble constant
and $\Omega_{\rm m0}$ represents the matter density parameter
at present.
Once $P_\kappa$ is known, 
one can straightforwardly convert it to other two-point statistics
such as the ellipticity correlation function \citep[e.g.,][]{2002A&A...396....1S}.

Note that the convergence power spectrum can be inferred 
directly through the cosmic shear field 
without resorting to the convergence field itself.
Thus, it can be measured without introducing any filter function.
The situation is the same for the convergence bispectrum.
This is in contrast to the peak counts and the MFs; 
one has to first construct
a convergence map with a filter before measuring them 
(see Sec.~\ref{sec:peak} for more detail).
This gives them an explicit dependence on the filter scale chosen for the map construction.
In what follows, the results should be interpreted with care as different statistics might probe different scales.
The scale is specified by the range of multipole moment $\ell$ 
for the two spectra, while it is given by the filter scale 
for peak counts and the MFs.

\subsubsection{Bispectrum}
\label{sec:bispec}

For the lensing convergence field, 
the bispectrum is defined as the three point correlation 
in Fourier space as
\beqa
\langle \tilde{\kappa}(\bm{\ell}_1) \tilde{\kappa}(\bm{\ell}_2) \tilde{\kappa}(\bm{\ell}_3)\rangle 
= (2\pi)^2 \delta_D (\bm{\ell}_{1}+\bm{\ell}_{2}+\bm{\ell}_{3})
B_\kappa (\bm{\ell}_{1}, \bm{\ell}_{2}, \bm{\ell}_{3}).
\eeqa
This quantity is zero for Gaussian fields and thus
$B_{\kappa}$ contains the lowest-order 
non-Gaussian information in the weak lensing field.
Similarly to the case of $P_{\kappa}$, one can relate the convergence bispectrum to 
the three-dimensional matter bispectrum $B_{\delta}$;
\beqa
B_{\kappa}(\bm{\ell}_{1}, \bm{\ell}_{2}, \bm{\ell}_{3})
= \int_{0}^{\chi_s} {\rm d}\chi \frac{W(\chi)^3}{r(\chi)^4} 
B_{\delta}\left(\bm{k}_{1}, \bm{k}_{2}, \bm{k}_{3}, z(\chi)\right)|_{\bm{k}_{i} = \bm{\ell}_{i}/\chi}.
\label{eq:kappa_bispec}
\eeqa
Recent studies have shown that 
the convergence bispectrum does supplement the power spectrum 
and we can gain $20-50$ percent in terms of
the signal-to-noise ratio (S/N)
up to a maximum multipole of a few thousands 
\citep[e.g.,][]{2013MNRAS.429..344K}.
However, the S/N from a combined analysis of 
the convergence power spectrum and bispectrum
is still significantly smaller than that of the ideal case of the Gaussian statistics.
This result motivates us to consider other statistical quantities
such as the peak counts and MFs.

\subsubsection{Peak counts}
\label{sec:peak}
The local maxima found in a smoothed convergence map
would have cosmological information originated from 
massive dark matter haloes 
and the superposition of large-scale structures
\citep[e.g.,][]{2004MNRAS.350..893H, 2010MNRAS.402.1049D, 2010PhRvD..81d3519K, 2011PhRvD..84d3529Y, 2016PASJ...68....4S}.
We here consider such local maxima and examine their statistical power in later sections.

In actual observations, one usually start with the cosmic shear
instead of the convergence field. 
The reconstruction of smoothed convergence is 
commonly based on the smoothed map of cosmic shear.
Let us first define the smoothed convergence map as
\beqa
{\cal K}(\bd{\theta}) = 
\int {\rm d}^2 \bd{\phi}\, \kappa(\bd{\theta}-\bd{\phi}) 
U (\bd{\phi}),
\eeqa
where $U$ is the filter function to be specified below.
We can calculate the same quantity by smoothing 
the shear field $\gamma$ as
\beqa
{\cal K} (\bd{\theta}) = \int {\rm d}^2 \bd{\phi} \ \gamma_{+}(\bd{\phi}:\bd{\theta}) Q_{+}(\bd{\phi}), \label{eq:ksm}
\eeqa
where $\gamma_{+}$ is the tangential component of the shear 
at position $\bd{\phi}$ relative to the point $\bd{\theta}$.
The filter function for the shear field $Q_{+}$ is related to $U$ by
\beqa
Q_{+}(\theta) = \int_{0}^{\theta} {\rm d}\theta^{\prime} \ \theta^{\prime} U(\theta^{\prime}) - U(\theta).
\label{eq:U_Q_fil}
\eeqa
We consider a filter function $Q_{+}$ that has a finite extent.
In such cases, one can write
\beqa
U(\theta) = 2\int_{\theta}^{\theta_{o}} {\rm d}\theta^{\prime} \ \frac{Q_{+}(\theta^{\prime})}{\theta^{\prime}} - Q_{+}(\theta),
\eeqa
where $\theta_{o}$ is the outer boundary of the filter function.

In the following, we consider the truncated Gaussian filter (for $U$):
\beqa
U(\theta) &=& \frac{1}{\pi \theta_{G}^{2}} \exp \left( -\frac{\theta^2}{\theta_{G}^2} \right)
\nonumber \\
&&\,\,\,\,\,\,\,\,\,\,
-\frac{1}{\pi \theta_{o}^2}\left[ 1-\exp \left(-\frac{\theta_{o}^2}{\theta_{G}^2} \right) \right], \\
Q_{+}(\theta) &=& \frac{1}{\pi \theta^{2}}\left[ 1-\left(1+\frac{\theta^2}{\theta_{G}^2}\right)\exp\left(-\frac{\theta^2}{\theta_{G}^2}\right)\right],
\label{eq:filter_gamma}
\eeqa
for $\theta \leq \theta_{o}$ and $U = Q_{+} = 0$ elsewhere.
Throughout this paper, we set $\theta_{o} = 10\times\theta_{G}$
and adopt $\theta_{G} = 1$ arcmin as a fiducial case.
Note that this choice of $\theta_{G}$ is 
considered to be an optimal smoothing scale for 
the detection of massive galaxy clusters
using weak lensing for $z_{\rm source}$ = 1.0 
\citep{2004MNRAS.350..893H}.

Let us now move to the peaks.
The height of peaks is in practice normalized as
$\nu(\bd{\theta})={\cal K}(\bd{\theta})/\sigma_{\rm shape}$,
where $\sigma_{\rm shape}$ is the noise variance coming from intrinsic ellipticity of galaxies.
We compute $\sigma_{\rm shape}$ following
\beqa
\sigma_{\rm shape}^2=\frac{\sigma_{\rm int}^2}{2n_{\rm gal}}\int_0^{\theta_{o}}{\rm d}\theta \, Q_{+}^2\left(\theta\right),
\label{eq.noise}
\eeqa
where $\sigma_{\rm int}$ is the rms value of the intrinsic ellipticity of 
the source galaxies and $n_{\rm gal}$ is the number density of galaxies. 
Unless otherwise stated, 
we assume $\sigma_{\rm int}=0.4$ 
and $n_{\rm gal}=10$ arcmin$^{-2}$ 
which are typical values for ground-based imaging surveys.

One can evaluate the smoothed convergence signal
arising from an isolated massive cluster at a given redshift
by assuming the matter density profile of dark matter halos \citep[e.g.,][]{1997ApJ...490..493N}.
Based on that, \citet{2004MNRAS.350..893H} present a simple theoretical framework 
to predict the number density of the peaks of the $\mathcal{K}$ field.
Their calculation provides a reasonable prediction 
when the S/N of $\nu$ 
due to massive halos is larger than $\sim4$
\citep[see,][for details]{2004MNRAS.350..893H}.
This is then refined by \citet{2010ApJ...719.1408F} by including the statistical properties 
of shape noise and its impact on the peak position.
We here focus on peak counts in a wider range of $\nu$ including peaks with low S/N, 
which is still difficult to predict with analytic approach.

\subsubsection{Minkowski functionals}
\label{sec:MFs}

MFs are morphological descriptors for smoothed random fields. 
There are three kinds of MFs
for two-dimensional maps. 
The functionals $V_0$, $V_1$ and $V_2$ represent
the area in which ${\cal K}$ is above the threshold ${\cal K}_{\rm thre}$, 
the total boundary length, the integral of geodesic curvature along the contours, respectively.
Hence, they are given by
\begin{eqnarray}
V_0({\cal K}_{\rm thre}) &\equiv& \frac{1}{A} \int_{Q} {\rm d}A, \\
V_1({\cal K}_{\rm thre}) &\equiv& \frac{1}{A} \int_{\partial Q} \frac{1}{4} {\rm d}\ell, \\
V_2({\cal K}_{\rm thre}) &\equiv& \frac{1}{A} \int_{\partial Q} \frac{1}{2\pi} K {\rm d}\ell,
\end{eqnarray}
where $K$ is the geodesic curvature of the contours, 
${\rm d}A$ and ${\rm d}\ell$ represent the area and length elements, 
and $A$ is the total area.
In the above, we also defined $Q$ and $\partial Q$, which are the excursion sets
and boundary sets for the smoothed field $\mathcal{K}(\bm{x})$, respectively.
They are given by
\begin{eqnarray}
Q = \{ \bm{x}|\mathcal{K} (\bm{x}) > {\cal K}_{\rm thre} \}, \\
\partial Q = \{ \bm{x}|\mathcal{K} (\bm{x}) = {\cal K}_{\rm thre} \}.
\end{eqnarray}
In particular, 
$V_2$ is equivalent to 
a kind of genus statistics and equal to the number of connected regions above the threshold, 
minus those below the threshold.
Therefore, for high thresholds, 
$V_2$ is almost the same as the peak counts.

For a two-dimensional Gaussian random field, 
the expectation values of MFs can be described as shown in \citet{1986PThPh..76..952T}:
\begin{eqnarray}
V_{0}({\cal K})&=&\frac{1}{2}\left[1-{\rm erf}
\left( \frac{{\cal K}-\bar{\cal K}}{\sigma}\right)\right],
\label{eq:v0_gauss} \\
V_{1}({\cal K})&=&\frac{1}{8\sqrt{2}}
\frac{\tau}{\sigma}\exp\left(-\frac{({\cal K}-\bar{\cal K})^2}{\sigma^2}\right),
\label{eq:v1_gauss} \\
V_{2}({\cal K})&=&
\frac{1}{2(2\pi)^{3/2}}
\left(\frac{{\cal K}-\bar{\cal K}}{\sigma}\right)
\frac{\tau^2}{\sigma^2} \nonumber \\
&&\,\,\,\,\,\,\,\,\,\,\,\,\,\,\,\,\,\,\,\,
\times\exp \left( -\frac{({\cal K}-\bar{\cal K})^2}{\sigma^2} \right),
\label{eq:v2_gauss}
\end{eqnarray}
where $\bar{\cal K}=\langle \mathcal{K} \rangle$, 
$\sigma^2 = \langle \mathcal{K}^2 \rangle - \bar{\cal K}^2$, and
$\tau^2 = \langle |\nabla \mathcal{K}|^2 \rangle$.
Although MFs can be evaluated perturbatively 
if the non-Gaussianity of the field is weak \citep{Matsubara2003,Matsubara2010},
it is difficult to adopt the perturbative approach 
for highly non-Gaussian fields \citep{Petri2013}.
In this paper, we pay a special attention to 
the non-Gaussian cosmological information 
obtained from convergence MFs.
Therefore, instead of analytical calculations, again, 
we consider the numerical measurements 
of MFs from the smoothed convergence field $\cal K$ estimated by Eq.~(\ref{eq:ksm}).
\citet{2015PhRvD..92f4024L} have demonstrated 
that lensing MFs can be a powerful probe of $f(R)$ gravity, 
while we will further investigate them with 
more detailed simulation of gravitational lensing in this paper.
The main difference between our analysis and \citet{2015PhRvD..92f4024L}
is in the method for the projection of the large-scale structure.
While our simulation properly takes into account the contribution from the structure along the line of sight 
by ray-tracing, 
\citet{2015PhRvD..92f4024L} have focused on the surface mass density field at a specific redshift of $\sim0.1$.

\section{SIMULATION AND ANALYSIS}\label{sec:sim_and_ana}
\subsection{$N$-body and ray-tracing simulations}
\label{subsec:sim}
We generate three-dimensional matter density fields
using a $N$-body code {\tt ECOSMOG} \citep{Li:2011vk},
which supports a wide class of modified gravity models including $f(R)$ gravity. 
This code is based on an adaptive mesh refinement code {\tt RAMSES}\footnote{\url{http://www.itp.uzh.ch/~teyssier/ramses/RAMSES.html}} 
\citep{2002A&A...385..337T}.
The simulation covers a comoving box length 
of $240\, h^{-1}$Mpc for each dimension, and the gravitational force 
is computed using a uniform $512^3$ root grid with seven levels of 
mesh refinement, corresponding to the maximum comoving 
spatial resolution of $3.6\, h^{-1}$kpc.
We proceed the mesh refinement when 
the effective particle number in a grid cell 
becomes larger than eight.
The density assignment and force interpolation are performed
with the triangular shaped cloud (TSC) kernel.
We generate the initial conditions using the parallel code 
{\tt mpgrafic}\footnote{\url{http://www2.iap.fr/users/pichon/mpgrafic.html}} 
developed by \citet{Prunet:2008fv}.
The initial redshift is set to $z_{\rm init}=85$, 
where we compute the linear matter transfer function using 
{\tt linger} \citep{1995astro.ph..6070B}.
As the fiducial cosmological model, 
we adopt the following cosmological parameters: 
the matter density parameter $\Omega_{\rm m0}=0.315$, 
the cosmological constant in units of the critical density $\Omega_{\Lambda 0}=0.685$, 
the amplitude of curvature perturbations 
$\ln(10^{10} A_{s})=3.089$ at $k=0.05 \, {\rm Mpc}^{-1}$,
the Hubble parameter $h=0.673$ and 
the scalar spectral index $n_s=0.945$.
These parameters are consistent with the result of
\citet{2015arXiv150201589P}.
For the HS model, 
we consider two variants with $|f_{\rm R0}|=10^{-5}$ and $10^{-6}$,
referred to F5 and F6, respectively.
We fix the initial density perturbations for these simulations and 
allow the amplitude of the current density fluctuations to vary 
among the models.
The mass variance within a sphere with a radius of $8\, {\rm Mpc}\, h^{-1}$
(denoted by $\sigma_{8}$) 
is therefore different in the three models; 0.830, 0.883, and 0.845 in $\Lambda$CDM, 
F5 and F6, respectively.
The cosmic shear statistics are known to be sensitive to the combination 
of $\Omega_{\rm m0}$ and $\sigma_{8}$ 
\citep[e.g., see][for a review]{2015RPPh...78h6901K}.
In order to study the degeneracy of the cosmological parameters and the modified gravity parameters,
we perform four additional sets of $\Lambda$CDM simulations
with different values of $\Omega_{\rm m0}$ and $\sigma_{8}$.
Table~\ref{tb:params} summarizes the parameters in our $N$-body simulations.

\begin{table*}
\caption{
	Cosmological parameters used for $N$-body simulations. 
	In addition to the parameter in $f(R)$ gravity,
	two parameters ($\Omega_\textrm{m}$, $\sigma_{8}$) 
	are changed by $\pm1\sigma$ of {\it Planck} 2015 constraint 
	\citep{2015arXiv150201589P}. 
	When we vary $\Omega_\textrm{m}$, we also change $\Omega_\Lambda$ 
	to keep the spatial flatness. 
	\label{tb:params}
	}
\begin{tabular}{@{}lcccccccl}
\hline
\hline
Run & $f_{\rm R0}$ & $\sigma_8$ & $\Omega_\textrm{m0}$ & 
$\Omega_\Lambda$ & No. of $N$-body sim. & No. of maps & Explanation \\ \hline
GR & 0 & 0.830 & 0.315 & 0.685 & 1 & 100 & 
Fiducial $\Lambda$CDM model \\
F5 & $-10^{-5}$ & 0.883 & 0.315 & 0.685 & 1 & 100 & 
HS model with $f_{\rm R0}=-10^{-5}$ \\
F6 & $-10^{-6}$ & 0.845 & 0.315 & 0.685 & 1 & 100 & 
HS model with $f_{\rm R0}=-10^{-6}$ \\
High $\Omega_\textrm{m0}$ & 0 & 0.830 & 0.335 & 0.665 & 1 & 100 & $1\sigma$ higher $\Omega_\textrm{m0}$ model \\
Low $\Omega_\textrm{m0}$ & 0 & 0.830 & 0.295 & 0.715 & 1 & 100 & $1\sigma$ lower $\Omega_\textrm{m0}$ model  \\
High $\sigma_{8}$ & 0 & 0.850 & 0.315 & 0.685 & 1 & 100 & 
$1\sigma$ higher $\sigma_{8}$ model\\
Low $\sigma_{8}$ & 0 & 0.810 & 0.315 & 0.685 & 1 & 100 & 
$1\sigma$ lower $\sigma_{8}$ model\\
\hline
\end{tabular}
\end{table*}

For ray-tracing simulations of gravitational lensing, 
we generate light cone outputs using multiple simulation boxes
in the following manner. 
Our simulation volumes are placed side-by-side to cover the past light cone of a hypothetical observer 
with an angular extent $5^{\circ}\times 5^{\circ}$, 
from $z=0$ to 1, similarly to the methods in  
\citet{2000ApJ...537....1W}, \citet{2001MNRAS.327..169H} and \citet{2009ApJ...701..945S}.
The exact configuration can be found in the last reference.
The angular grid size of our maps is 
$5^{\circ}/4096\sim 0.075$ arcmin.
For a given cosmological model,
we use constant-time snapshots stored at various redshifts.
We create multiple light-cones out of these snapshots by randomly shifting
the simulation boxes in order to avoid the same structure appearing 
multiple times along a line of sight.
In total, we generate 100 quasi-independent lensing maps 
with the source redshift of $z_{\rm source}=1$
from our $N$-body simulation.
See \citet{2016PhRvD..93f3524P} for the validity of recycling
one $N$-body simulation to have multiple weak-lensing maps.

Throughout this paper,
we include galaxy shape noise $e$ in our simulation by
adding to the measured shear signal random ellipticities
which follow the two-dimensional Gaussian distribution
as
\beqa
P(e)=\frac{1}{\pi\sigma_{\rm e}^2}
\exp\left(-\frac{e^2}{\sigma_{\rm e}^2}\right),
\eeqa
where 
$e = \sqrt{e_{1}^2 + e_{2}^2}$ 
and
$\sigma^2_{\rm e} = \sigma_{\rm int}^2/(n_{\rm gal}\theta^2_{\rm pix})$ with 
the pixel size of $\theta_{\rm pix}=0.075$ arcmin.

\subsection{Statistical analyses}
\label{subsec:ana}
In the following, we summarize our methods to
measure cosmic shear statistics of interest 
from simulated lensing field.

\subsubsection*{Power spectrum}

We follow the method in \citet{2009ApJ...701..945S} 
to estimate the convergence power spectrum 
from numerical simulations based on the fast Fourier transform.
Namely, we measure the binned power spectrum of the convergence field 
by averaging the product of Fourier modes 
$|\tilde{\kappa}(\bd{\ell})|^2$
obtained by two-dimensional fast fourier transform.
We employ 30 bins logarithmically spaced in the range of 
$\ell = 100$ to $5\times10^{4}$. 
However, we consider 10 bins on $\ell<2,000$ in evaluating the 
expected signal level on modified gravity, since smaller scales
are in general more difficult to predict without theoretical uncertainties,
such as baryonic physics \citep[e.g.][]{2013PhRvD..87d3509Z,
2015ApJ...806..186O} or 
intrinsic alignment \citep[for a review, see e.g.][]{2015PhR...558....1T}.


\subsubsection*{Bispectrum}

We follow the method in 
\citet{2012A&A...541A.161V} and \citet{2013PhRvD..87l3538S} 
to estimate the convergence bispectrum, which is 
a straightforward extension of 
the power spectrum measurement.
We measure the binned bispectrum of the convergence field 
by averaging the product of three Fourier modes 
${\rm Re}[\tilde{\kappa}(\bd{\ell}_1)
\tilde{\kappa}(\bd{\ell}_2)\tilde{\kappa}(\bd{\ell}_3)]$
where ${\rm Re}[\cdots]$ represents the real part of a complex number.
We use 12 bins logarithmically spaced in the range of 
$\ell_{i} (i=1, 2, 3) = 100$ to $1\times10^{4}$ for each of the three multipoles, and focus
on bins in which all the multipoles are less than $2,000$ in later sections for the same reason as the power spectrum.

\subsubsection*{Peak counts}

We identify convergence peaks as follows.
Starting from the discretized $\kappa$ fields given on grid
obtained from numerical simulations,
we apply the filter function (\ref{eq:U_Q_fil}) to have a smooth
field $\mathcal{K}$. We then
define the peak as a pixel which has a higher value than all of its 
eight neighbour pixels.
We then measure the number of peaks 
as a function of $\mathcal{K}$.
We exclude the region within 2$\theta_{G}$ 
from the boundary of map 
in order to avoid the effect of incomplete smoothing.
These procedures are similar to the method in \citet{2015PhRvD..91f3507L}.
We consider 18 bins in the range of 
$-4 <\nu < 7$. However, we exclude bins with $\nu>4$
in the discussion of the statistical power since
we recycle one simulation to obtain multiple convergence maps
and massive halos corresponding to such high peaks
are heavily affected by the cosmic variance in that one
realization\footnote{
Although the abundance of these high peaks in our 
simulations is broadly explained 
by a simple analytical model~\citep{2016MNRAS.459.2762H}, 
more quantitative analyses are required to assess
any systematic effects on the abundance of high peaks.
}.

\begin{figure*}
\centering
\includegraphics[width=0.90\columnwidth, bb=0 0 468 442]
{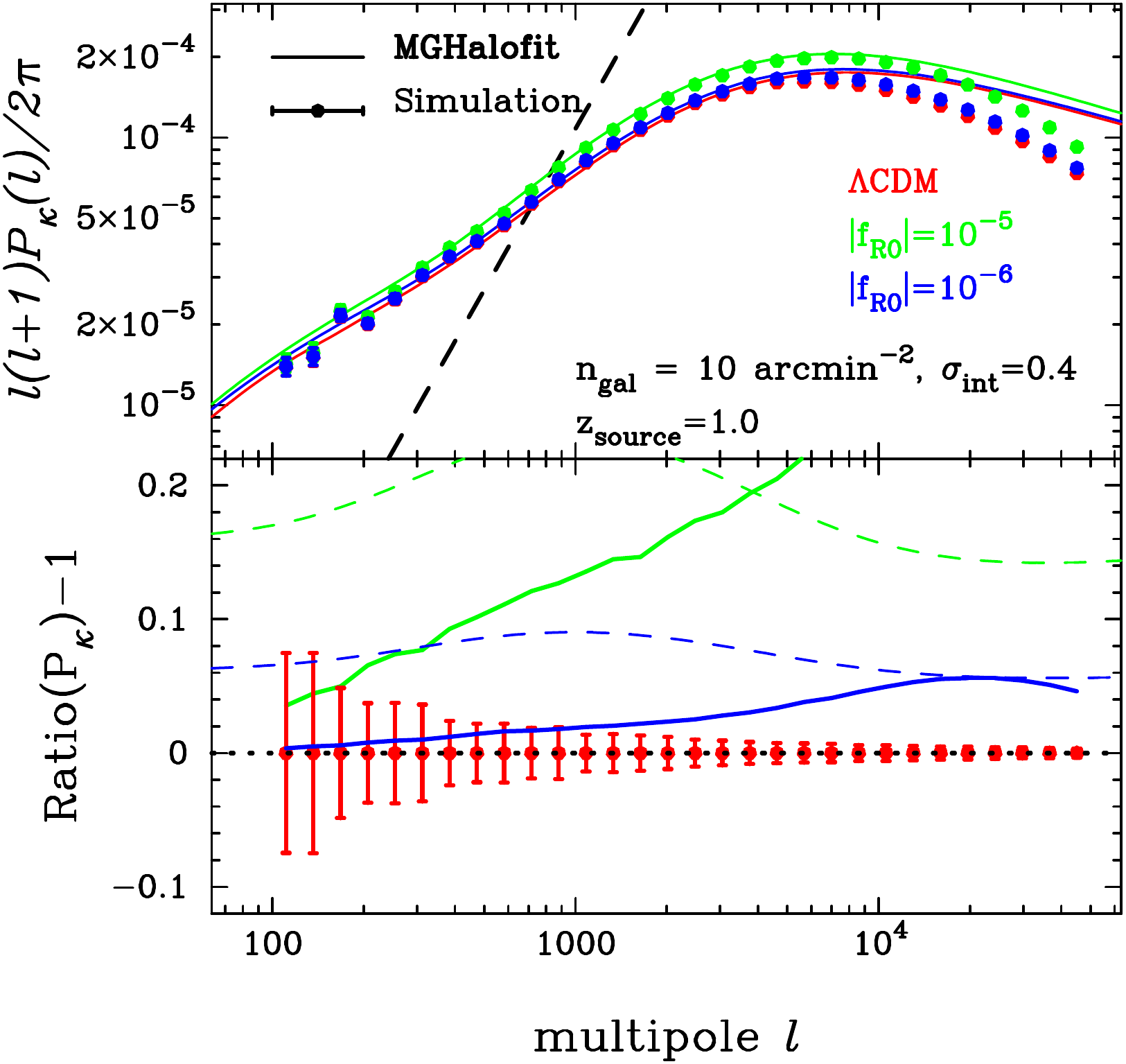}
\includegraphics[width=0.84\columnwidth, bb=0 0 437 486]
{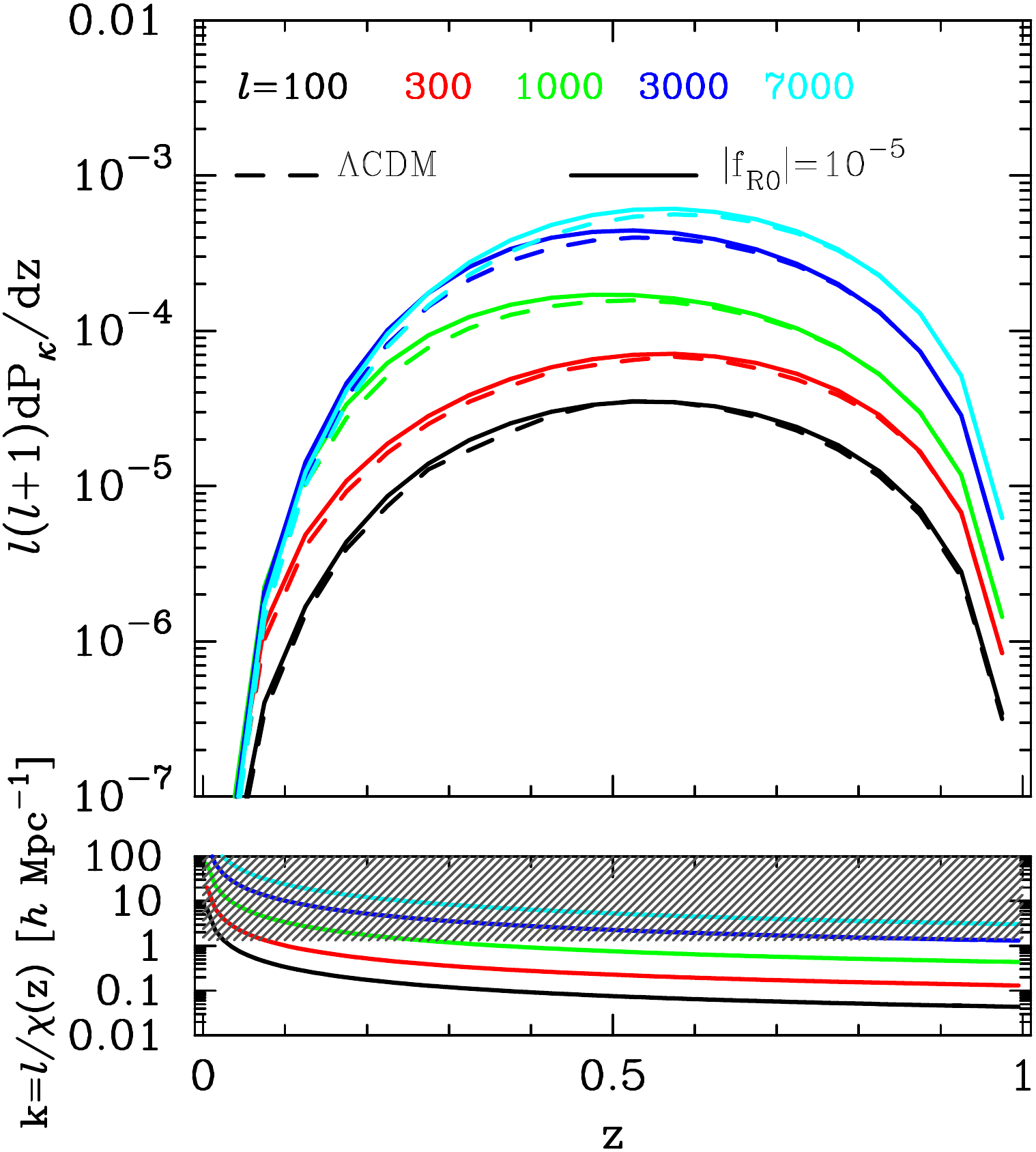}
\caption{
	Impact of $f(R)$ gravity on the convergence power spectrum.
	{\it Left}: we show the dependence on $|f_{\rm R0}|$
	of the convergence power spectrum.
	In the top panel, the colored points represent
	the average power spectrum over 100 realizations
	for the three models, while 
	the bottom shows the relative difference   
	between $\Lambda$CDM and the two $f(R)$ models.
	The black dashed line in the top panel corresponds to 
	the shape noise contribution, while the colored lines are 
	theoretical models based 
	on a fitting formula of the three-dimensional matter
	power spectrum \citep{2014ApJS..211...23Z}.
	In the bottom panel, colored dashed line represents
	the relative difference of the convergence power spectrum 
	for $\Lambda$CDM model when we vary the value of $\sigma_{8}$
	to match to those in the $f(R)$ models.
	In the left-hand panels, the error bars represent the standard error 
	of the average (i.e. the standard deviation 
	of the \textit{each} measurement divided by $\sqrt{100}$).
	{\it Right}: We show the integrand of the convergence power spectrum
	(\ref{eq:kappa_power}) as a function of redshift $z$. 
	In the top panel, 
	the dashed lines correspond to the $\Lambda$CDM case, 
	while the solid lines are for the F5 model. 
	There, different colored lines show the case of different multipoles 	as shown in the figure legend.
	In the lower panel, we show the comoving scale $k$ that contributes 
	to the convergence power spectrum at the multipole $\ell$
	at a given redshift $z$.
	As a reference, the grey hatched region represents
	the region where the linear matter perturbations would 
	be enhanced by the additional scalar field
	degree of freedom.
	}
\label{fig:kappa_power}
\end{figure*} 

\subsubsection*{Minkowski functionals}

After constructing the smooth convergence field $\mathcal{K}$ on grid exactly as
in the peak counts, we apply
the following estimators of MFs, 
as shown in, e.g., \citet{Kratochvil2012},
\begin{eqnarray}
V_0({\cal K}_{\rm thre}) &=& \frac{1}{A} \int 
\Theta (\mathcal{K}-{\cal K}_{\rm thre})
{\rm d}x{\rm d}y, \\
V_1({\cal K}_{\rm thre}) &=& \frac{1}{4A} \int 
\delta_D (\mathcal{K}-{\cal K}_{\rm thre}) \sqrt{\mathcal{K}_x^2+\mathcal{K}_y^2} 
{\rm d}x{\rm d}y, \\
V_2({\cal K}_{\rm thre}) &=& \frac{1}{2\pi A} \int 
\delta_D (\mathcal{K}-{\cal K}_{\rm thre}) \nonumber \\
&\times&
\frac{2\mathcal{K}_x\mathcal{K}_y\mathcal{K}_{xy}-\mathcal{K}_x^2\mathcal{K}_{yy}-\mathcal{K}_y^2\mathcal{K}_{xx}}{\mathcal{K}_x^2+\mathcal{K}_y^2} {\rm d}x{\rm d}y,
\end{eqnarray}
where $\Theta(x)$ is the Heaviside step function and 
$\delta_D(x)$ is the Dirac delta function. 
The integrals in the above expressions are carried out by summing up the 
values over grid points specified by the sky coordinates $x$ and $y$.
The subscripts on ${\cal K}$ represent differentiation with respect to 
these coordinates. 
The first and second differentiation are evaluated with finite difference.
We compute MFs for 100 equally spaced bins of 
$(\mathcal{K}-\langle \mathcal{K} \rangle)/\sigma$ 
between $-10$ and $10$. We consider only the range
$-3<(\mathcal{K}-\langle \mathcal{K} \rangle)/\sigma<4$ in
the detectability analysis, for similar reasons to the peak counts.
We will see shortly that a large amount of the sensitivity to 
the parameter $|f_{R0}|$ comes from this range.

\section{RESULTS}
\label{sec:res}

\subsection{Dependence of parameter in $f(R)$ gravity}

\subsubsection*{Power spectrum}

Let us first show the result of the convergence power spectrum 
$P_{\kappa}$.
The left-hand panels in figure~\ref{fig:kappa_power} summarize
the average convergence power spectrum 
obtained from 100 ray-tracing maps.
In both top and bottom panels,
the red, green and blue points (or lines) correspond to
the $\Lambda$CDM, F5 and F6 model, respectively.
The red, green and blue solid lines in the top panel 
represent the corresponding theoretical predictions based on Eq.~(\ref{eq:kappa_power}).
To calculate Eq.~(\ref{eq:kappa_power}) for the $f(R)$ models,
we adopt the fitting formula of three-dimensional matter power spectrum
as developed in \citet{2014ApJS..211...23Z}.
Note that this fitting formula can 
reproduce the result in \citet{2012ApJ...761..152T} 
in the case of $|f_{\rm R0}|=0$.
We find that 
the predication provides a reasonable fit to
our simulation results for three different models in the range of 
$\ell \simlt 7000$.
In the bottom panel, we show the relative difference of $P_{\kappa}$
between the $f(R)$ models and $\Lambda$CDM.
The red error bars in the bottom panel corresponds to the standard error of the 
average convergence power spectrum for $\Lambda$CDM model.
We confirm that the F5 and F6 models 
change the convergence power spectrum 
in the range of $\ell<7000$ by $\simlt$20\% 
and $\simlt4\%$, respectively.
In comparison, we also consider the relative difference of $P_{\kappa}$ 
between two $\Lambda$CDM models with different values 
of $\sigma_{8}$.
The green dashed line in the bottom left-hand panel shows the relative 
difference between $\sigma_{8}=0.883$ and $0.830$,
while the blue one is for the difference between 
$\sigma_{8}=0.845$ and $0.830$ 
(see also Table~\ref{tb:params}).
While the overall level of the enhancement of power is similar to the modified gravity simulations, 
the trend in the dashed lines is 
quite different from that in the solid lines.
Therefore, the convergence power spectrum 
can be a useful probe
of $f(R)$ gravity 
whereas the effect is partly compensated by a change of $\sigma_{8}$.

\begin{figure*}
\centering
\includegraphics[width=0.86\columnwidth, bb=0 0 436 427]
{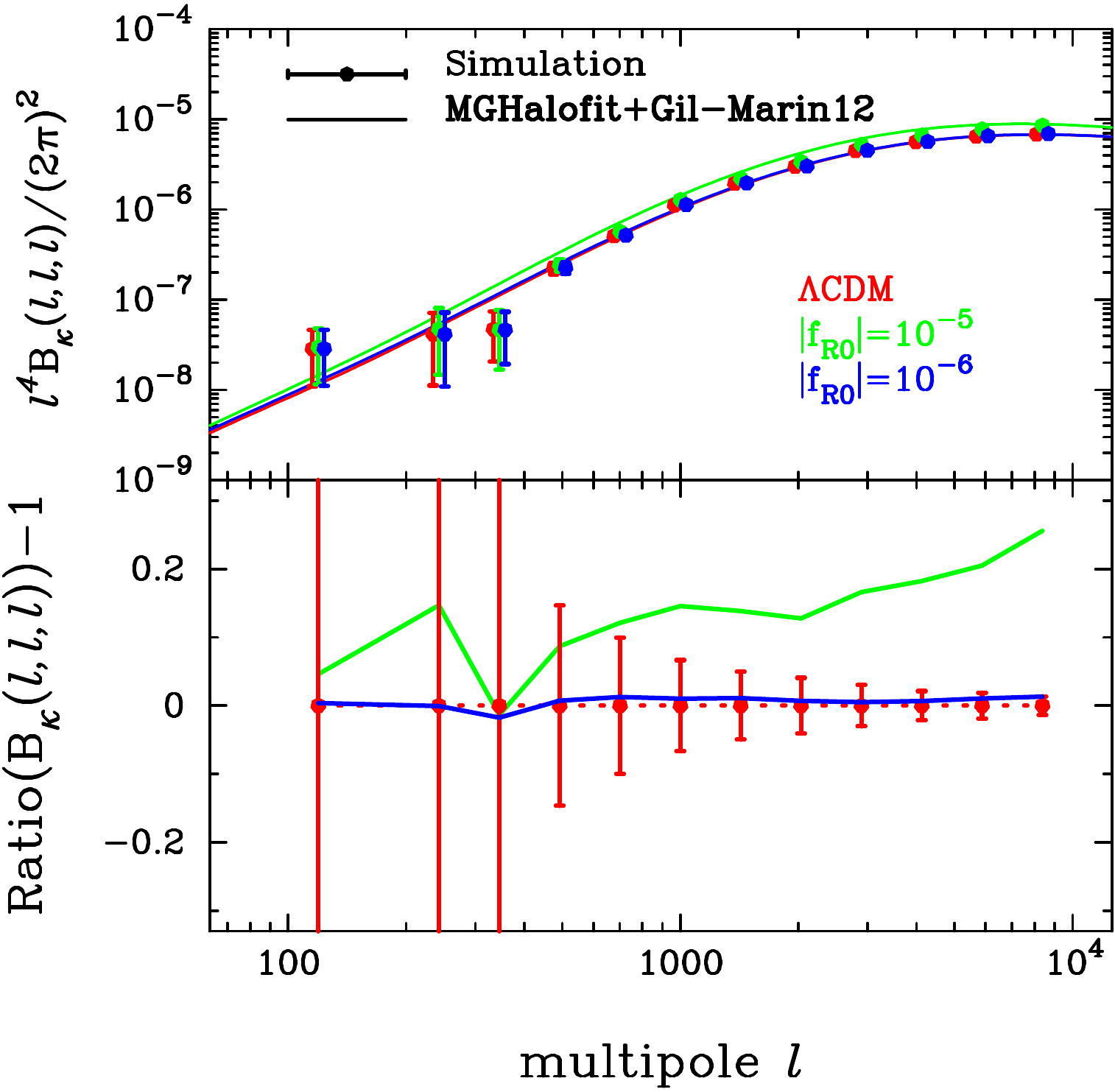}
\includegraphics[width=0.90\columnwidth, bb=0 0 494 470]
{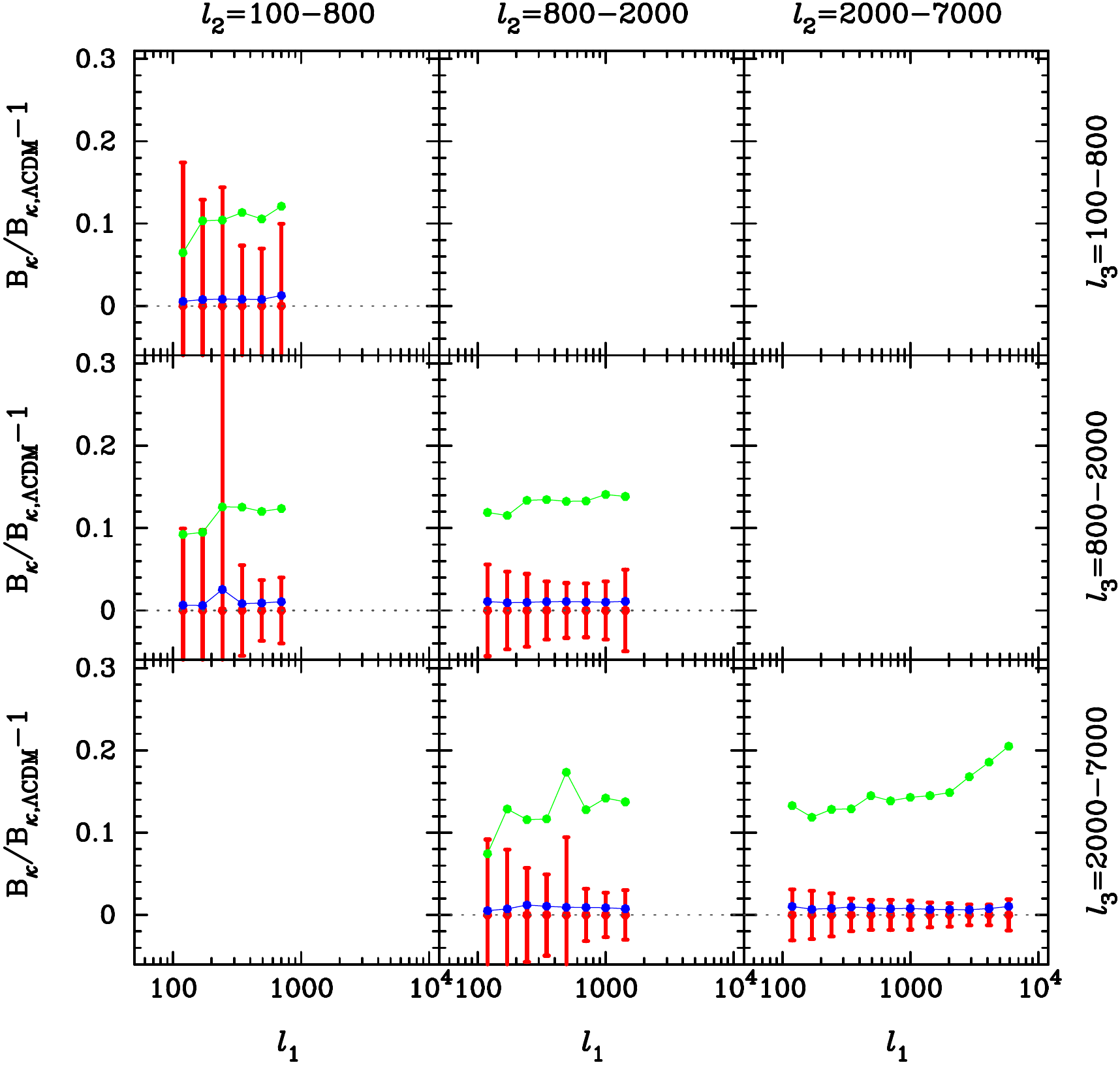}
\caption{
	Impact of $f(R)$ gravity on the convergence bispectrum.
	{\it Left}: We show the dependence on $|f_{\rm R0}|$
	of the convergence bispectrum for equilateral configuration 
	with $\ell_1 = \ell_2 = \ell_3 = \ell$.
	In the top panel, colored points represent
	the average bispectrum over 100 ray-tracing realizations
	for three models, while 
	the bottom left one shows the relative difference 
	between $\Lambda$CDM and the two $f(R)$ models.
	The colored solid line in the top panel shows 
	the theoretical model based on
	Eq.~(\ref{eq:kappa_bispec}) with the fitting formula 
	in \citet{2012JCAP...02..047G}.
	The error bars represent the standard error 
	on the estimated average (i.e. the standard deviation divided by $\sqrt{100}$).
	{\it Right}: The relative difference of the convergence bispectrum 
	between $\Lambda$CDM and the $f(R)$ models for 
	more general triangular configurations $(\ell_1, \ell_2, \ell_3)$
	as shown on the axes.
	The red error bars show 
	the standard error on the average bispectrum
	for $\Lambda$CDM model, while the green and blue lines 
	are the ratio for F5 and F6 model, respectively.
	Note that we impose the condition 
	of $\ell_{1} \le \ell_{2} \le \ell_3$
	to count every triangle configuration once.
	}
\label{fig:kappa_bispec}
\end{figure*} 

We further examine the contribution to $P_{\kappa}$
from the lens at a given redshift to understand when $f(R)$ gravity 
enhances the projected power the most significantly.
The top right-hand panel in figure~\ref{fig:kappa_power} shows the integrand 
in Eq.~(\ref{eq:kappa_power}) using the fitting formula 
in \citet{2014ApJS..211...23Z}.
Compared to $\Lambda$CDM, F5 model enhances the amplitude 
of the matter density fluctuations at $z\simlt0.6$ for all the multipoles depicted here. 
In the bottom right panel, we show the wavenumber 
$k(z) = \ell/\chi(z)$
contributing
in the calculation of Eq.~(\ref{eq:kappa_power}) for a given redshift $z$.
On linear scales where $\delta \rho_{m} \ll \bar{\rho}_{m}$,
the Compton wavelength of the extra scalar field $f_{R}$ can be
expressed as
\beqa
\lambda_{C}^{-1} = \left(\frac{1}{3(n+1)}
\frac{\bar{R}}{|f_{\rm R0}|} \left(\frac{\bar{R}}{\bar{R}_0}\right)^{n+1}
\right)^{1/2}.
\eeqa
The grey hatched region in the bottom panel 
represents $k > a\lambda_{C}^{-1}$ 
where the fifth force due to $f_{R}$ can efficiently enhance 
the linear density fluctuations.
Although the competition between the non-linear gravitational growth and the chameleon 
mechanism would make the situation more complicated, 
the criterion of $k > a\lambda_{C}^{-1}$ provides
the typical scale where $f(R)$ gravity affects the density distribution.
On large scales where $\ell \simlt 300$,
the linear approximation works fairly well 
and the deviation from $\Lambda$CDM can be mainly explained 
by the scale-dependent linear growth rate.
On the other hand, 
the chameleon mechanism does not 
completely suppress the effect of $f(R)$ gravity 
on the matter distribution on small scales.

\subsubsection*{Bispectrum}

We next consider the convergence bispectrum $B_{\kappa}$.
Figure~\ref{fig:kappa_bispec} summarizes the simulation results 
obtained from 100 ray-tracing maps for the three different models
with $|f_{\rm R0}|=0, 10^{-6}$ and $10^{-5}$.
In the left-hand panels, we show the result of $B_{\kappa}$ for
the equilateral triangle configuration with $\ell_1 = \ell_2 = \ell_3 = \ell$.
First of all, we compare the simulation result for the $\Lambda$CDM model 
and the theoretical prediction. 
In the calculation, 
we adopt the fitting formula of the
three-dimensional matter bispectrum $B_{\delta}$
proposed in \citet{2012JCAP...02..047G} and plug it into Eq.~(\ref{eq:kappa_bispec}).
This fitting formula explicitly includes the three-dimensional
matter power spectrum and we use the fitting formula in \citet{2014ApJS..211...23Z} 
(which is equivalent to \citep{2012ApJ...761..152T} for $\Lambda$CDM) 
for that.
We find that the fitting formula is in good agreement
with the simulation results again over the range of $\ell \simlt 7000$
for $\Lambda$CDM model.
This result is consistent with a previous work 
by \citet{2013PhRvD..87l3538S}. 
For $\ell\simlt300$, the difference between the simulation results and
the analytic models appears to be relatively large, mildly larger than
the error bars estimated from the scatter among realizations.
This is possibly because we repeatedly use a single $N$-body simulation
to generate a quasi-independent ensemble and thus the quoted error level
might be not very accurate. Another reason for the small discrepancy is
the finite area of the simulated maps. This can be expressed as
a convolution with a window function corresponding to the $5^{\circ}\times5^{\circ}$ 
geometry, and this corresponds to $\ell = 360/5 = 72$ in multipole.
The low-$\ell$ data points are not so far from this typical length scale.
Furthermore, the fitting formula can also provide a reasonable fit
to both F5 and F6 models, even though 
the fitting formula for $B_{\delta}$ is constructed 
for a $\Lambda$CDM cosmology by numerical simulations.
In order to quantify the effect of $|f_{\rm R0}|$ on $B_{\kappa}$,
we also show the relative difference of the bispectrum 
between the $\Lambda$CDM and the $f(R)$ models 
in the left-hand bottom panel 
and the right-hand panels of Figure~\ref{fig:kappa_bispec}.
The bottom left-hand panel represents the result 
for the equilateral configuration, while the right-hand panels summarize
more general configurations specified by three multipoles, $\ell_{1}\le \ell_{2} \le \ell_{3}$.
In the right panels, we reduce the number of bins 
for $\ell_{2}$ and $\ell_{3}$ 
to show the effect of $|f_{\rm R0}|$ in an easy to see manner.
Overall, we find that the F5 model affects the convergence bispectrum
by $\simlt20\%$ and 
the 
dependence on the triangle shape 
is rather weak except for $\ell_{i}\simgt2000$.
On the other hand, we can not find significant deviation from the $\Lambda$CDM for F6 model.
Although the effect of $|f_{\rm R0}|$ on $B_{\kappa}$
seems similar to that on $P_{\kappa}$ for the angular scale 
of $\ell\simlt2000$, 
the statistical uncertainty of $B_{\kappa}$ would be larger than
$P_{\kappa}$, 
implying that the bispectrum would be less sensitive to $f(R)$ gravity
and provide a weaker constraint on $|f_{\rm R0}|$ 
compared to the power spectrum.
We revisit the constraining power 
on $|f_{\rm R0}|$ with cosmic shear statistics
in Section~\ref{subsec:detectability}.
Nevertheless, we should note that $B_{\kappa}$ would play an important
role to break the degeneracy among cosmological parameters such
as $\Omega_{\rm m0}$ and $\sigma_{8}$ in cosmic shear analyses.

\subsubsection*{Peak count}

\begin{figure*}
\centering
\includegraphics[width=0.86\columnwidth,bb=0 0 471 491]
{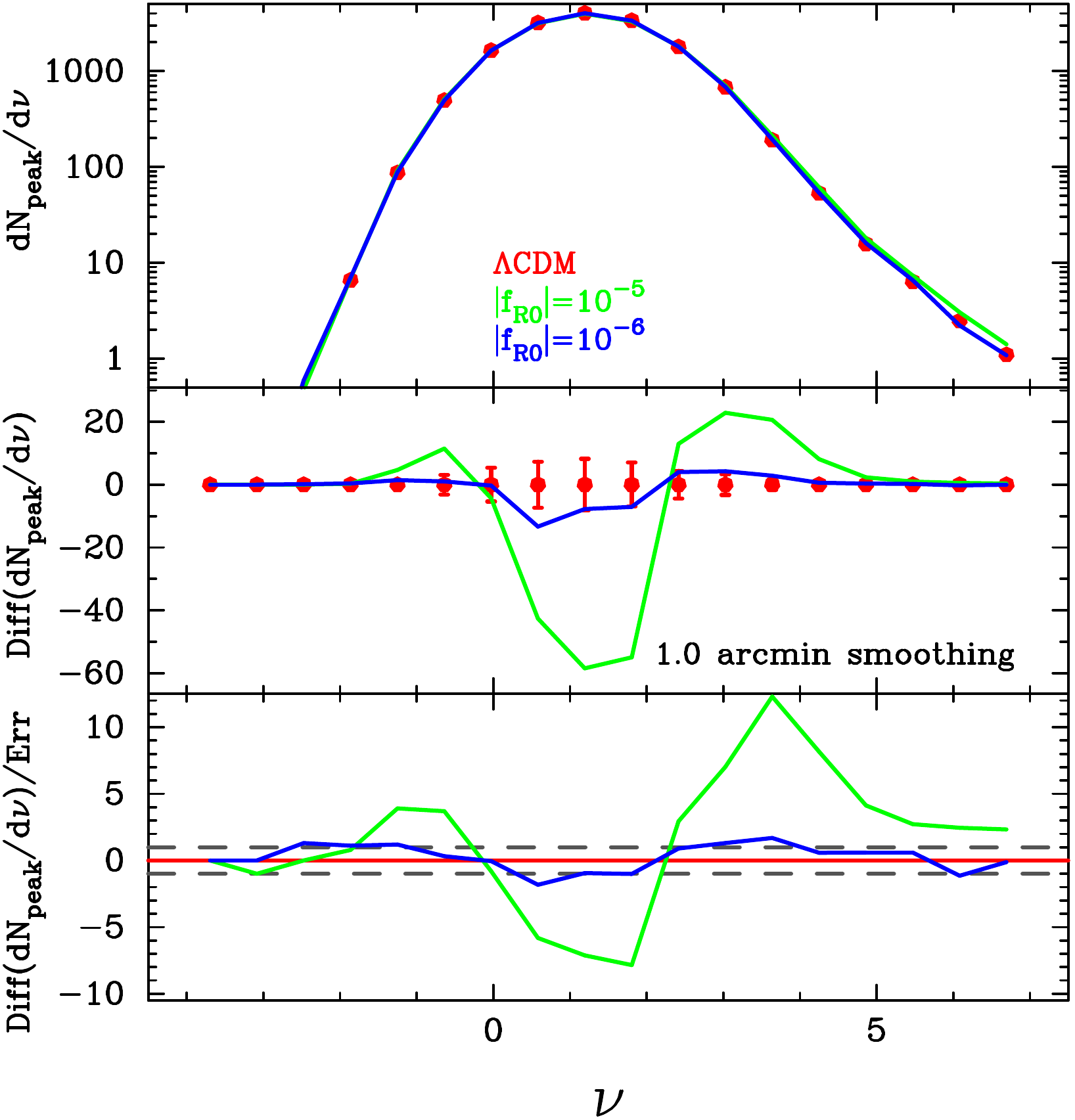}
\includegraphics[width=0.86\columnwidth,bb=0 0 471 491]
{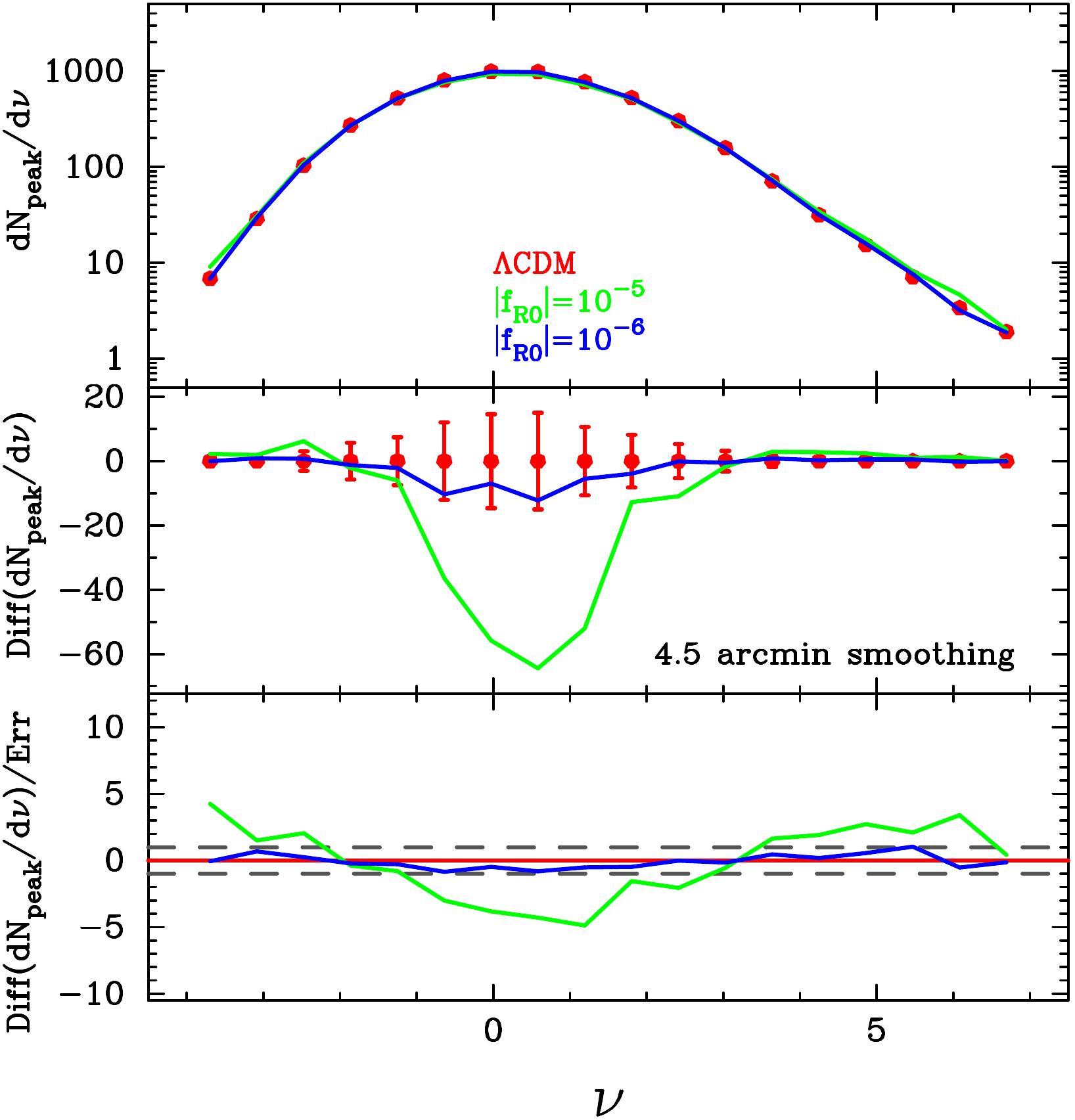}
\caption{
	Impact of $f(R)$ gravity on the peak counts.
	{\it Left}: 
	We show the dependence of the peak counts on $|f_{\rm R0}|$.
	In the top panel, the red points represent
	the average (differential) number density of peaks 
	over 100 realizations for the $\Lambda$CDM model, while
	the green and blue lines are for the F5 and F6 models. 
	The middle one shows the difference  
	between $\Lambda$CDM and $f(R)$ models,
	while the bottom is for the difference normalized 
	by the standard error of average for the $\Lambda$CDM model.
	{\it Right}: Similar to the left panel, 
	but larger smoothing scale of 4.5 arcmin is adopted.
	In both panels, the error bars represent the standard error 
	of the average (i.e. the standard deviation divided by $\sqrt{100}$).
	Note that the error level corresponds to
	$100\times{(5\,\mathrm{deg})^2} = 2,500\,\mathrm{deg}^2$.
	}
\label{fig:kappa_peaks}
\end{figure*} 

We here summarize the results of the peak counts.
We define the differential number density of peaks
and then compare the results among three different models.
Figure~\ref{fig:kappa_peaks} shows the effect of $f(R)$ gravity
on the peak counts.
The left-hand panel represents the simulation results with the smoothing scale 
of 1 arcmin, while the right-hand panel are for the smoothing with 
$\theta_G=4.5$ arcmin.
In both panels, red, green and blue points (or lines) represent
the average of number density of peaks for $\Lambda$CDM,
F5, and F6 models, respectively.
As in Figure~\ref{fig:kappa_power},
we show the difference of the number density in the middle panels,
while we normalize the difference by the standard error of average 
for the $\Lambda$CDM model in the bottom panels.
We find that the effect of $f(R)$ gravity on the peak counts 
appears in not only $\nu \ge 3$ but also $\nu \sim 1$.
The peaks with $\nu \ge 3$ correspond to isolated 
massive dark matter halos along the line of sight 
[see \citet{2010ApJ...719.1408F} for analytical estimate
of the shape-noise contamination on these peaks and also
\citet{2016MNRAS.459.2762H} 
for the detailed comparisons in $f(R)$ model].
General trend of the number density among three models 
is found to be consistent with the expectation from 
the halo mass function \citep[e.g.,][]{2016PASJ...68....4S}.
The number density of high peaks increases in the range of $\nu\ge3$ in the HS model.
These specific features would reflect the non-trivial dependence 
of halo mass function on $|f_{\rm R0}|$ 
\citep[e.g.,][]{2011PhRvD..84h4033L, 2012MNRAS.421.1431L,
2013PhRvD..87l3511L, 2016arXiv160708788C}.
With a larger smoothing scale, which
roughly corresponds to the removal of Fourier modes with 
$\ell\simgt2000$, 
a bumpy feature at $\nu\sim3.5$ for the F5 model disappears.
For larger $\theta_G$, the halo-peak correspondence gets worse
because sharp structures such as halos 
are erased by the smoothing operation.
This would indicate that the simple framework presented in \citet{2015MNRAS.453.3043S} can not explain 
the number count of peaks on $\nu>3$ as $\theta_G$ would
become larger.
Also, we find the number density at $\nu\sim 1$
is significantly changed from $\Lambda$CDM 
when we set $|f_{\rm R0}|=10^{-5}$.
This could originate from the larger density fluctuations 
in F5 and F6 models expressed in terms of $\sigma_{8}$ \citep{2010PhRvD..81d3519K}
or the superposition of less massive objects 
\citep{2011PhRvD..84d3529Y}.
Although it is still not straightforward to physically interpret these
low-S/N peaks and thus it might be safer to avoid them 
in cosmological tests, we would like to stress here that they do have 
sensitivity to the parameter $|f_{\rm R0}|$ in a \textit{statistical} sense, 
even in the presence of realistic shape noise.
We reserve the study on the degeneracy between $|f_{\rm R0}|$
and $\sigma_{8}$ in peak counts in Section~\ref{subsec:degeneracy}.

\subsubsection*{Minkowski functionals}

\begin{figure*}
\centering
\includegraphics[width=0.90\columnwidth,bb=0 0 508 469]
{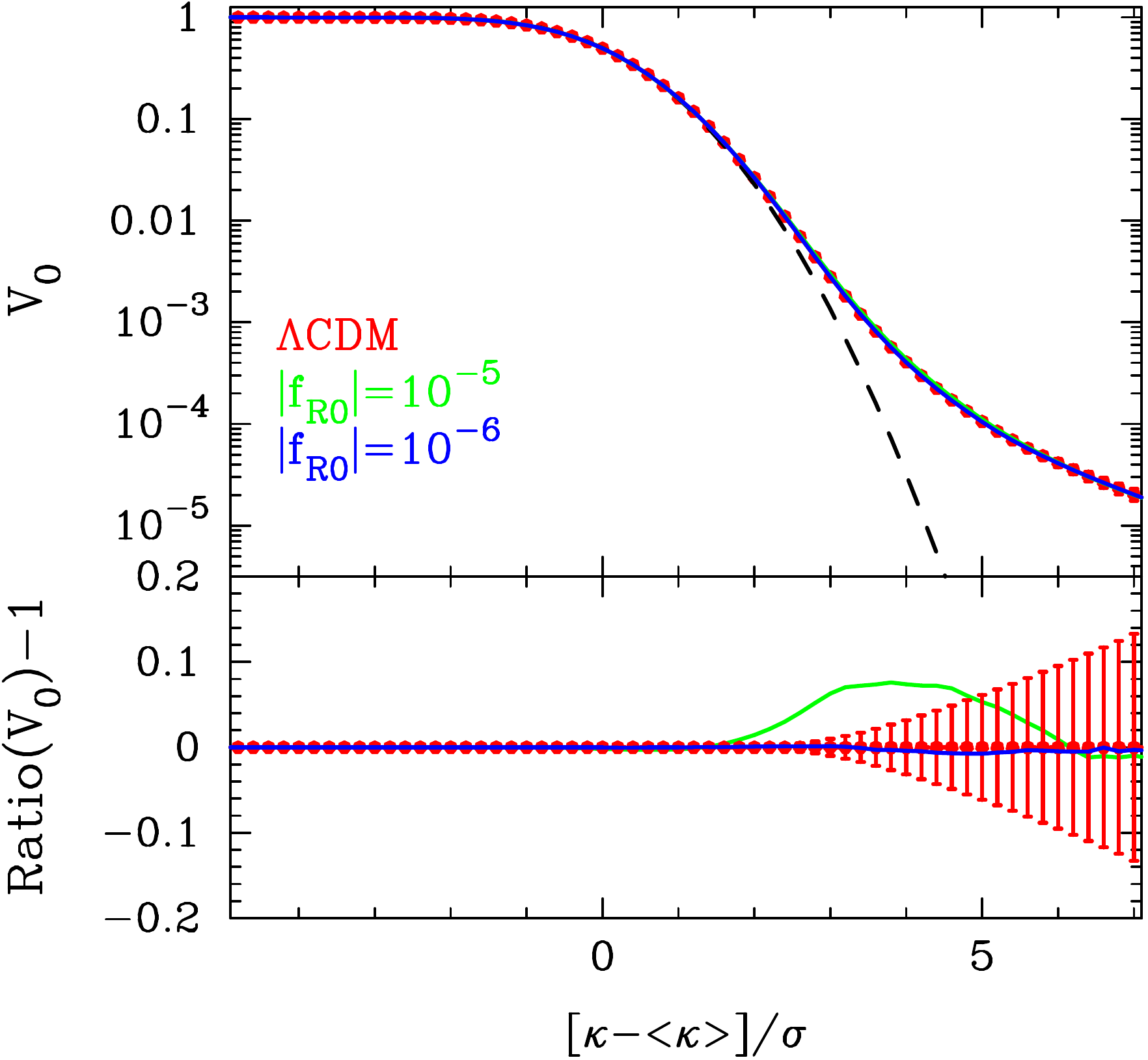}
\includegraphics[width=0.90\columnwidth,bb=0 0 508 467]
{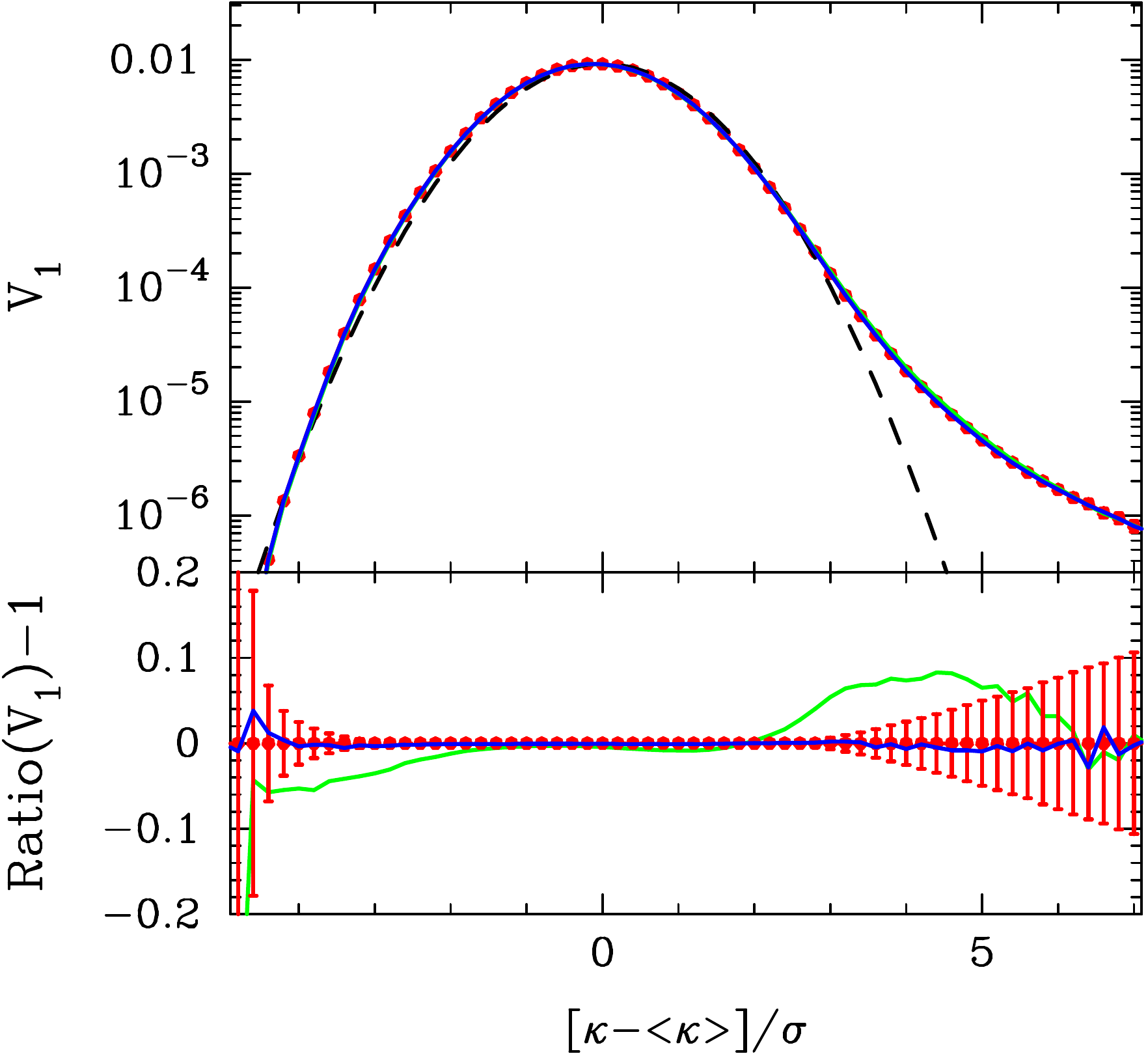}
\includegraphics[width=0.90\columnwidth,bb=0 0 508 467]
{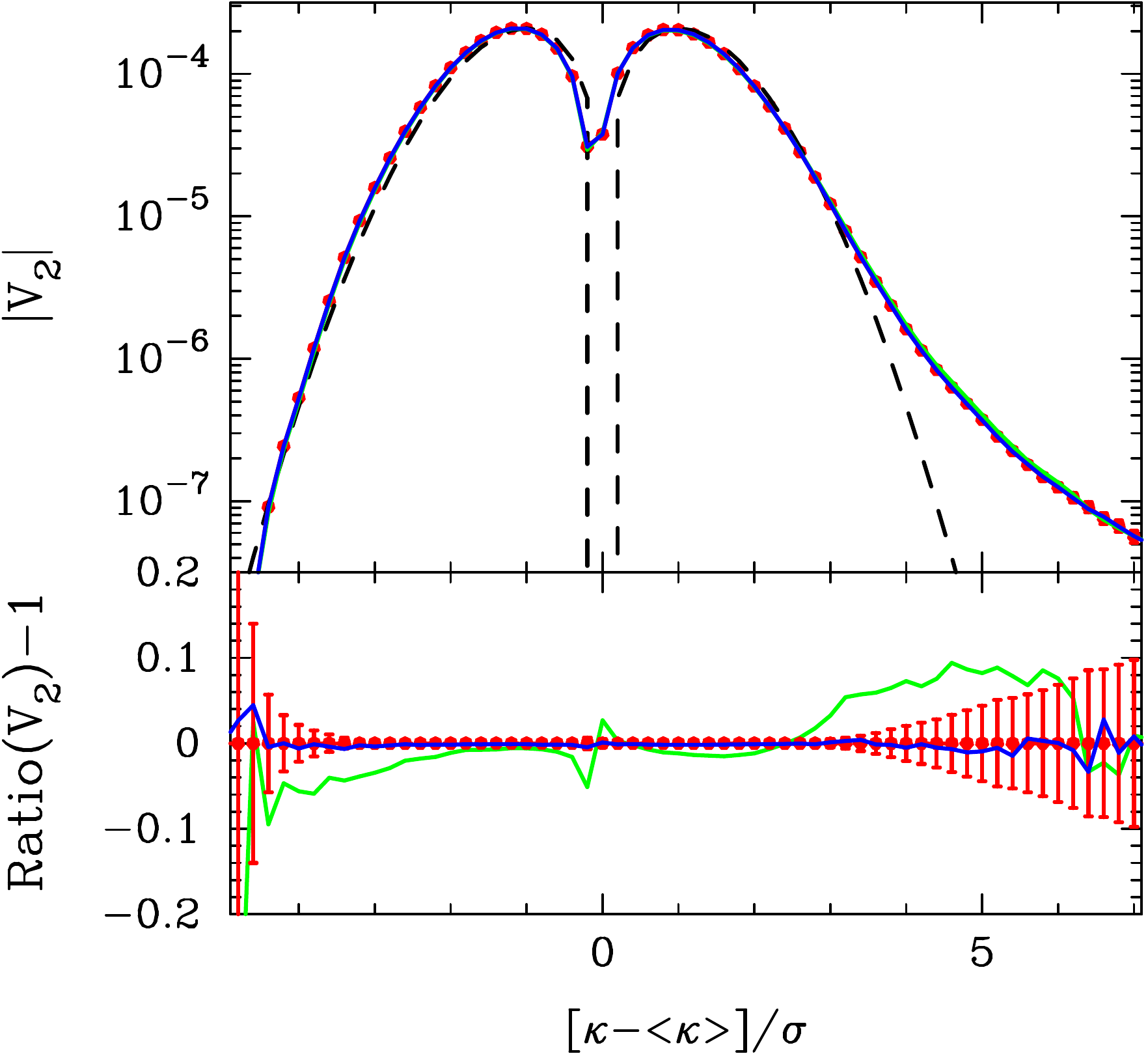}
\caption{
	Impact of $f(R)$ gravity on the Minkowski functionals (MFs).
	The three panels represent the results of 
	$V_{0}$, $V_{1}$ and $V_{2}$.
	In every panel, the error bars represent the standard error 
	of the average (i.e. the standard deviation divided by $\sqrt{100}$).
	Also, colored points represent
	the average MF over 100 realizations for the three models, while 
	the bottom portion shows the relative difference of MF 
	between $\Lambda$CDM and $f(R)$ models.
	The dashed line shows the Gaussian prediction 
	for $\Lambda$CDM model.
	}
\label{fig:kappa_MFs}
\end{figure*} 

We then present the measurements of the lensing MFs
obtained from 100 simulations.
Figure~\ref{fig:kappa_MFs} summarizes the effect 
of $|f_{\rm R0}|$ on the lensing MFs.
First of all, we confirm 
the non-Gaussian feature in lensing MFs for the three models
even when we add the shape noise for which we assume Gaussian distribution.
The shape of the MFs obtained from simulations 
can not be explained by the Gaussian expectation in Eqs.~(\ref{eq:v0_gauss})-(\ref{eq:v2_gauss}) depicted by the dashed lines,
implying that the lensing MFs 
are useful probe of non-Gaussian nature of the convergence field that can not be captured by the power spectrum.
Our results are broadly consistent with
a previous work by \citet{2015PhRvD..92f4024L}.
In the case of F5 model, 
we find that the deviation from $\Lambda$CDM 
is at most $\sim10\%$ and the clear deviations 
are found at $x = ({\cal K}-\langle {\cal K}\rangle)/\sigma \sim 2-5$.
On the other hand, we find only $\simlt1\%$ differences 
between F6 and $\Lambda$CDM models.
Note that the deviation from the $\Lambda$CDM we observe 
is found to be smaller than \citet{2015PhRvD..92f4024L} have shown.
One of the reasons behind this trend is in the difference of the adopted 
values of $|f_{\rm R0}|$ and other cosmological parameters.
The model parameters used in \citet{2015PhRvD..92f4024L} are 
different despite the same label: 
their F5 means $|f_{\rm R0}|=1.29\times10^{-5}$ instead of $10^{-5}$.
They also adopted smaller $\Omega_{\rm m0}$ and $\sigma_{8}$, 
both indicating weaker screening and 
therefore stronger deviation from GR for the same $|f_{\rm R0}|$.
Besides, the difference between our result and previous one
would be partly explained by the projection effect. 
Our simulations include the projection effect and the shape noise
simultaneously, while \citet{2015PhRvD..92f4024L} have focused 
on the surface mass density at $z=0.1$.
Furthermore, we find that 
the difference of the lensing MFs between the $\Lambda$CDM and the HS model has the similar trend to a change of $\sigma_{8}$
\citep[e.g., see Figure 2 in][]{2014ApJ...786...43S}.


\subsection{Detectability of imprint of $f(R)$ gravity}
\label{subsec:detectability}

In order to quantify the detectability of $f(R)$ gravity in a
given statistical quantity, 
we start by writing a measure of a goodness-of-fit;
\beqa
\chi^2 &=& 
\sum_{i,j}
{\bf C}^{-1}_{ij} 
\left[{\cal O}(x_{i}; true)-{\cal M}(x_{i}; test)\right] \nonumber \\
&&\,\,\,\,\,\,\,\,\,\,\,\,\,\,\,\,\,\,\,\,\,\,\,\,\,
\times
\left[{\cal O}(x_{j}; true)-{\cal M}(x_{j}; test)\right],
\label{eq:dist_chi_sq}
\eeqa
where ${\cal M}(x_{i}; test)$ 
represents a theoretical model of 
cosmic shear statistic 
at the $i$-th bin of $x$ for a cosmological model that one wishes to test, 
${\cal O}(x_{i}; true)$ is an observed statistic drawn from the true unknown
cosmology, and 
${\bf C}$ is the covariance matrix of the observed data vector ${\cal O}$.
In our case, ${\cal O}$ 
corresponds to either the power spectrum, bispectrum,
peak counts or MFs, while $x$ refers to 
the multipole $\ell$, the peak height
$\nu$, or the normalized convergence 
$({\cal K}-\langle {\cal K}\rangle)/\sigma$ 
depending on the statistics.
In what follows, we also consider a data vector 
${\cal O}$ composed of different statistics 
when we examine parameter
constraints from joint analyses of more than one statistic. 
In such cases, we properly take into account 
the off-diagonal components relevant to the two statistics 
of interest in the covariance matrix.

When ${\cal O}$ follows a multi-variate Gaussian distribution
{\it and} {\it if} we assume the correct model in $\mathcal{M}(x_i;test)$,
the quantity defined by Eq.~(\ref{eq:dist_chi_sq}) 
follows the $\chi^2$ distribution with the degree of freedom
of $N_{\rm bin}-1$ as the name suggests, 
where $N_{\rm bin}$ represents the total number of 
bins for the observables ${\cal O}$.
Borrowing the idea behind Eq.~(\ref{eq:dist_chi_sq}), which compares the
levels of estimated (in the form of a covariance matrix)
and measured (the actual scatter around the mean) 
cosmic variances, we define a similar quantity to assess the
statistical power to constrain $|f_{\rm R0}|$ by replacing the
numerator with
the difference of the expected statistics in two models:
\beqa
\left({\rm S/N}\right)^2 &=& 
\sum_{i,j}
{\bf C}^{-1}_{ij} 
\left[{\cal M}(x_{i}; |f_{\rm R0}|)
-{\cal M}(x_{i}; \Lambda{\rm CDM})\right] \nonumber \\
&&\,\,\,\,\,\,\,\,\,\,\,
\times
\left[{\cal M}(x_{j}; |f_{\rm R0}|)
-{\cal M}(x_{j}; \Lambda{\rm CDM})\right],
\label{eq:dist_chi_sq_test}
\eeqa
where we consider the
$f(R)$ cosmology characterized by $|f_{\rm R0}|$ and the fiducial 
$\Lambda$CDM cosmology.

One can assess the discriminating power of the statistic by the $({\rm S/N})^2$ defined above.
Note that this quantity does not depend on the binning scheme explicitly,
as long as we take a binning fine enough not to miss important features in the 
statistics. 
In the absence of degeneracy between 
different cosmological parameters, 
one can straightforwardly convert the $({\rm S/N})^2$ 
to the expected level of constraint on $|f_{R0}|$;
$({\rm S/N})^2=4$ for $|f_{R0}|=10^{-5}$ corresponds to $\sigma_{|f_{R0}|}=10^{-5}/\sqrt{4}=5\times10^{-6}$, for instance.
Note that we consider the degeneracy among cosmological parameters
in Section~\ref{subsec:degeneracy} in details.

In Eq~(\ref{eq:dist_chi_sq_test}),
we estimate ${\cal M}(x_{i}; |f_{\rm R0}|)$ and 
${\cal M}(x_{i}; \Lambda{\rm CDM})$ as the ensemble average 
over 100 realizations of our ray-tracing maps
as shown in Section~\ref{subsec:sim}.
To derive accurate covariance matrices of these observables
and the cross covariance between two different observables,
we also make use of the 1000 ray-tracing simulations performed 
by \citet{2009ApJ...701..945S}.
The maps in \citet{2009ApJ...701..945S} 
have almost the same design as our simulations, 
but are generated for slightly different cosmological parameters 
[consistent with Wilkinson Microwave Anisotropy Probe (WMAP) 3-year results
\citep{2007ApJS..170..377S}].
We use the maps with a sky coverage of $5\times5\, {\rm deg}^{2}$
for $z_{\rm source}=1$.
In order to estimate cosmic shear statistics, 
we properly take into account
the contamination from the intrinsic shape of sources by adding 
a Gaussian noise to shear (also see Section~\ref{subsec:sim}).
For a given observable ${\cal O}$, 
we estimate 
the covariance matrix using the 1000 realizations 
of ray-tracing simulations
as follows;
\beqa
{\bf C}_{ij} &=& 
\frac{1}{N_{\rm rea}-1}
\sum_{r=1}^{N_{\rm rea}}
\left[{\cal O}^{(r)}(x_{i})-\bar{{\cal O}}(x_{i})\right] \nonumber \\
&&\,\,\,\,\,\,\,\,\,\,\,\,\,\,\,
\,\,\,\,\,\,\,\,\,\,\,\,\,\,\,\,
\,\,\,\,\,\,\,\,\,\,\,\,\,\,
\times
\left[{\cal O}^{(r)}(x_{j})-\bar{{\cal O}}(x_{j})\right], \\
\bar{{\cal O}}(x_{i}) &=& 
\frac{1}{N_{\rm rea}}
\sum_{r=1}^{N_{\rm rea}}{\cal O}^{(r)}(x_{i}),
\eeqa
where $N_{\rm rea}=1000$ and 
${\cal O}^{(r)}(x_{i})$ is the observable obtained from 
$r$-th realization of simulations for $i$-th bin of $x$.
When we calculate the inverse covariance matrix, 
we multiply a debiasing correction,
$\alpha=(N_{\rm rea}-N_{\rm bin}-2)/(N_{\rm rea}-1)$,
with $N_{\rm rea}=1000$ 
and $N_{\rm bin}$ being the number of total bins 
in our data vector \citep{2007A&A...464..399H}.
In the following, we assume that the covariance matrix is scaled as the
inverse of the survey area and consider a hypothetical lensing survey 
with a sky coverage of 
$1500\, {\rm deg}^{2}$, which corresponds to 
the ongoing imaging survey with 
Subaru Hyper Suprime-Cam\footnote{\rm{http://www.naoj.org/Projects/HSC/j\_index.html}}\citep{2006SPIE.6269E...9M}.
Note that under the assumed scaling of the covariance matrix, one can easily calculate 
the corresponding $({\rm S/N})^2$ for a given sky coverage of $A\, {\rm deg}^{2}$
by multiplying the $({\rm S/N})^2$ presented in this section by a factor of $A/1,500$.
Table~\ref{tb:chi_sq} summarizes the results of this section.

\begin{table*}
\caption{
	Summary of the signal-to-noise ratio among cosmic shear statistics.
	We show $({\rm S/N})^2$ defined by Eq.~(\ref{eq:dist_chi_sq}).
	We show the results for $|f_{R0}|=10^{-5}$ in the upper half, and for $|f_{R0}|=10^{-6}$ in the bottom half. 
	In addition to the fiducial analysis, we also show the results when we increase the smoothing scale 
	to $\theta_{G} = 4.5\, {\rm arcmin}$, or the source number density is set to 
	$n_{\rm gal}=30\, {\rm arcmin}^{-2})$.
	All values are for the sky coverage of 1,500 squared degrees.
	\label{tb:chi_sq}
	}
\begin{tabular}{@{}lcccccccl}
\hline
\hline
 & $P_{\kappa}$ & $B_{\kappa}$ & Peak & MFs & $P_{\kappa}+B_{\kappa}$
 & $P_{\kappa}+$Peak & $P_{\kappa}+$MFs 
 \\ \hline
$|f_{\rm R0}|=10^{-5}$ \\ \hline 
Fiducial analysis 
 & 127.6 & 45.1 & 121.7 & 1000.9 & 130.5 & 201.6 & 1066.0
 \\ 
Larger smoothing scale 
 & 127.6 & 45.1 & 53.5 & 35.3 & 130.5 & 151.1 & 185.1
 \\
Higher source number density
 & 255.1 & 73.9 & 608.2 & 2837.2 & 264.5 & 668.6 & 2972.2
 \\ \hline
$|f_{\rm R0}|=10^{-6}$ \\ \hline 
Fiducial analysis 
 & 2.63 & 0.301 & 4.85 & 4.28 & 2.64 & 6.52 & 6.51
 \\ 
Larger smoothing scale
 & 2.63 & 0.301 & 1.39 & 0.237 & 2.64 & 3.37 & 3.45
 \\
Higher source number density 
 & 5.44 & 0.540 & 29.0 & 19.3 & 6.49 & 30.3 & 27.3
 \\
\hline 
\end{tabular}
\end{table*}

\subsubsection*{Fiducial analysis}

As the fiducial analysis, 
we consider a situation in which the two spectra $P_{\kappa}$ and $B_{\kappa}$
are measured in the range of $100\le\ell\le2000$,
while we have the number density of peaks in the range of $-2 < \nu < 4$
and MFs are given for $-3\le({\cal K}-\langle {\cal K}\rangle)/\sigma \le 4$.
For the two convergence spectra, we adopt the same binning as summarized in 
Section~\ref{subsec:ana}, leading to 14 bins for $P_{\kappa}$
and 78 bins for $B_{\kappa}$ in total.
On the other hand, we construct the number density of peaks 
in 10 bins with width of $\Delta \nu = 0.6$, while 
we employ 12 bins to measure each lensing MF.

As shown in Table~\ref{tb:chi_sq},
the values of $({\rm S/N})^2$ indicates that all the cosmic shear statistics considered here
can distinguish the F5 model from the $\Lambda$CDM model
with a high significance level when the other cosmological parameters are fully known.
Even a very small modification to GR such as our F6 model is detectable
by these statistics except when we employ the bispectrum alone.
Surprisingly, non-standard statistics such as the peak counts or the MFs
have very high signal-to-noise ratio competitive or significantly larger than
the conventional analyses using the power or bispectra \citep{2016PhRvL.117e1101L}.
This indicates that the weak lensing convergence field indeed exhibits 
strong non-Gaussianity that is difficult to capture by low-order 
polyspectra.
Geometrical measures such as the MFs are especially powerful in such a regime.
We also find the $({\rm S/N})^2$ from combined analyses with two statistics
is 
slightly smaller than the simple sum of the individual values (the three right-hand columns).
This is due to the cross covariance between the statistics.
While we can access independent information through different measures
reflecting an increase in the $({\rm S/N})^2$ from the combined analyses, 
part of it is common to that in the power spectrum.
This demonstrates the importance of a proper account of the 
cross-covariance in actual data analyses.



\subsubsection*{Dependence on smoothing scale and shape noise}

We have seen so far the statistical power of four different measures in testing the possibility of modified gravity.
Among the four statistics, the power and the bispectra are given explicitly as a function of the physical scale.
Since we expect that smaller scales are more severely contaminated 
by e.g., intrinsic alignment or baryonic physics, 
we can conduct a more reliable cosmological test by limiting ourselves in large scales.
On the other hand, the dependence of the peak counts and the MFs on the physical scale is less clear.
What we have done so far is based on the statistics at one given scale $\theta_{G} = 1\, {\rm arcmin}$, chosen to 
have a good correspondence between peaks and massive clusters.
We thus investigate in this section the dependence of the detectability of a non-zero $|f_{R0}|$ on the smoothing scale
that defines the peaks and MFs. We also test the dependence on the mean density of source galaxies, 
which can alter the
result significantly.

We first examine the dependence on the smoothing scales by 
setting $\theta_{G}=4.5\, {\rm arcmin}$, which
roughly corresponds to $\ell = 2000$.
For this choice of smoothing, all the four statistics 
probe roughly the same angular scale and thus we expect that they have a similar level of theoretical uncertainties due to small scale effects.
In this sense, we can do a fairer comparison among the four.
The results are shown in Table~\ref{tb:chi_sq} both for $|f_{R0}|=10^{-5}$ and $10^{-6}$.
Compared to the fiducial analysis with $\theta_G=1\,{\rm arcmin}$, the level of non-Gaussianity in the smoothed map is strongly suppressed.
As a result, the S/N from the lensing MFs is greatly reduced. 
The statistical power of the peak counts is also suppressed, but to a much lesser extent.
When we combine these statistics with the power spectrum, 
we have a larger increase in the S/N for the MFs than for peak counts,
because there is a larger overlap in information for the peak counts 
and the power spectrum with this choice of $\theta_G$.
Although we lose significant $({\rm S/N})^2$ in the MFs, they still probe independent information to the power spectrum.

Another test is a larger source number density with the smoothing scale unchanged. 
We consider 30 ${\rm arcmin}^{-2}$, and this leads to the shape noise level reduced by a factor of $1/\sqrt{3}\sim0.57$.
We reanalyse the power and the bispectra in addition to the peak counts and the MFs in this case.
We can see in Table~\ref{tb:chi_sq} that such a deeper imaging survey with a higher source number density
provides a significantly improved S/N. 
Especially, the peak counts and lensing MFs are sensitive 
to the number density.
We find that $({\rm S/N})^2$ for peak counts and MFs
increases by a factor of $\sim5$ and $\sim2.8$, respectively.
Provided that the small scale uncertainties are well under control, these two statistics have a potential
to distinguish the F5 or even F6 model from the $\Lambda$CDM model with a very high significance
in upcoming deep imaging surveys.

\subsection{Degeneracy among cosmological parameters}
\label{subsec:degeneracy}

We then study the degeneracy between $|f_{\rm R0}|$
and cosmological parameters.
Cosmic shear observables depend sensitively on 
the two parameters of $\Omega_{\rm m0}$ and $\sigma_{8}$
through Eq.~(\ref{eq:kappa_delta}).
Thus, we focus on these parameters and investigate 
how well we can constrain $|f_{\rm R0}|$ when these parameters
are jointly varied.
We first construct a model of the parameter dependence of 
cosmic shear statistic ${\cal O}(x)$ by expanding into the 
Taylor series around a fiducial point 
$(\Omega_{\rm m0, fid}, \sigma_{8, {\rm fid}}, |f_{\rm R0}|=0)$:
\beqa
{\cal O}(x_{i}; \Omega_{\rm m0}, \sigma_{8}, |f_{\rm R0}|) 
&\simeq& {\cal O}(x_{i}; \Omega_{\rm m0, fid}, \sigma_{8, {\rm fid}}, |f_{\rm R0}|=0) \nonumber \\
&&
+ \frac{\partial {\cal O}(x_{i})}{\partial \Omega_{\rm m0}} 
(\Omega_{\rm m0}-\Omega_{\rm m0, fid}) \nonumber \\
&&
+ \frac{\partial {\cal O}(x_{i})}{\partial \sigma_{8}}
(\sigma_{8}-\sigma_{8, {\rm fid}}) \nonumber \\
&&
+ \frac{\partial {\cal O}(x_{i})}{\partial |f_{\rm R0}|}
|f_{\rm R0}|,
\label{eq:model_obs_LCDM}
\eeqa
where 
$\Omega_{\rm m0, fid}=0.315$, 
$\sigma_{8, {\rm fid}}=0.830$,
and
the first derivatives are
estimated from the five simulations at the bottom of
Table~\ref{tb:params} 
by the finite difference method
(both-sided for $\Omega_{\rm m0}$ and $\sigma_8$,
and one-sided for $|f_{\rm R0}|$).

\begin{figure*}
\centering
\includegraphics[width=1.5\columnwidth, bb=0 0 705 429]
{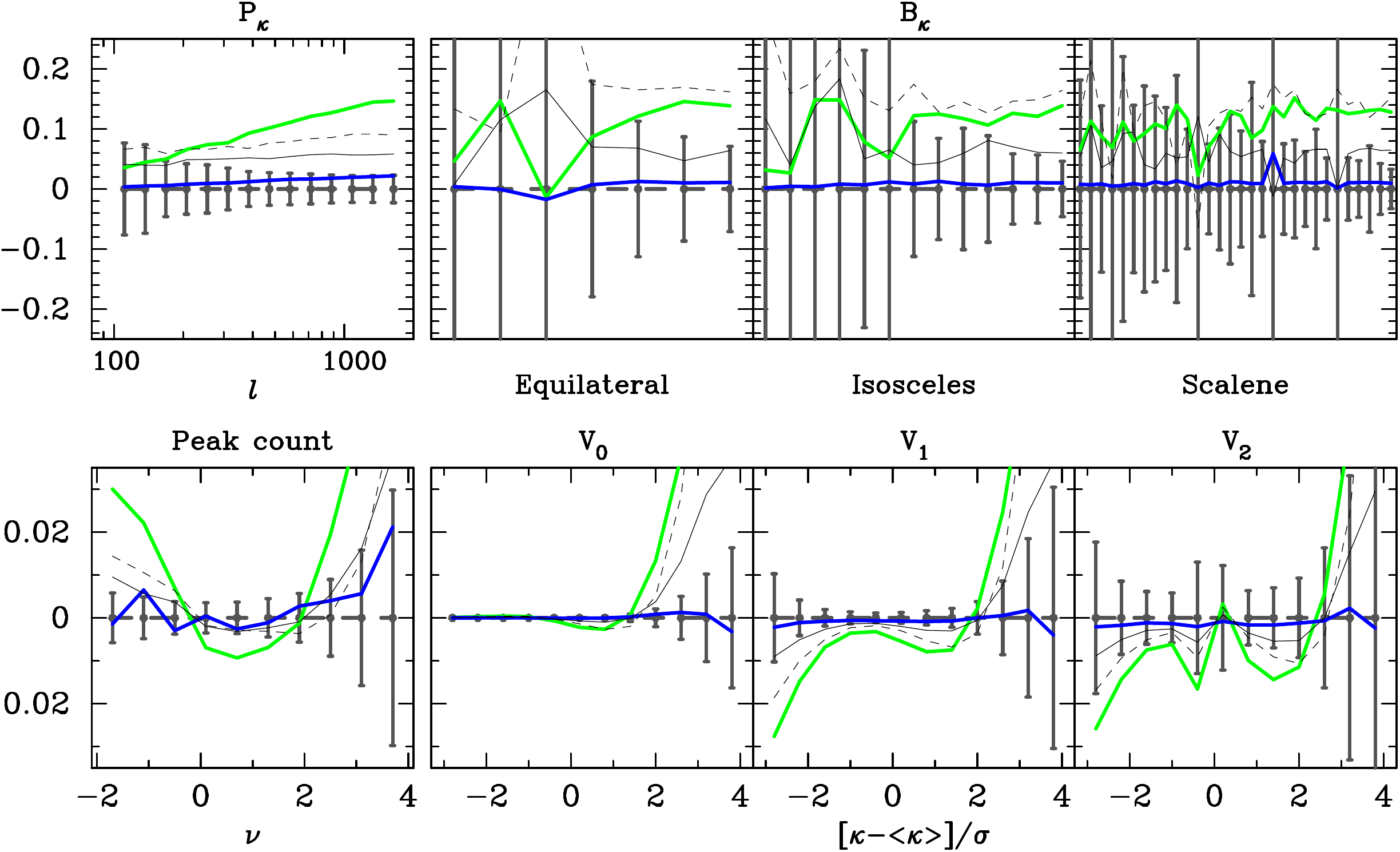}
\caption{
	Degeneracy between cosmological parameters and $f(R)$ gravity.
	Each panel represents the relative difference 
	of one of a various cosmic shear statistics
	between the fiducial $\Lambda$CDM model and the other models.
	The green and blue lines correspond to the F5 and F6 model, respectively.
	On the other hand, the black solid line shows the case of 
	$\Lambda$ CDM with 
	$\Omega_{\rm m0}$ larger by 0.025, 
	while the black dashed line is for a model with $\sigma_{8}$ larger by 0.050.
	Note that the grey error bars in each panel correspond to 
	the cosmic variance from a survey with the sky coverage of 1,500 squared degrees.
	The detailed configuration of triangles in bispectrum 
	is summarized in Appendix.
	}
\label{fig:degen_stat}
\end{figure*}

Figure~\ref{fig:degen_stat} summarizes our cosmic shear statistics
as a function of $|f_{\rm R0}|$, $\Omega_{\rm m0}$ 
and $\sigma_{8}$.
In this section, we consider the same binning of observables 
as in the fiducial analysis 
in Section~\ref{subsec:detectability}.
Figure~\ref{fig:degen_stat} shows 
the fractional difference of cosmic shear statistic compared to the fiducial model. 
The green and blue lines in the figure represent the ratio for F5 and F6 model, respectively.
The black solid and dashed lines are for $\Lambda$CDM 
with higher $\Omega_{\rm m0}$ and $\sigma_{8}$.
For visualization, we classify the triangular configuration of the arguments of the convergence bispectrum 
into equilateral $(\ell_1=\ell_2=\ell_3)$,
isosceles $(\ell_1 = \ell_2)$, 
and scalene $(\ell_1 \neq \ell_2 \neq \ell_3)$.
The grey error bars represent the statistical uncertainty 
in a hypothetical survey with 1,500 squared degrees estimated 
from the 1000 ray-tracing simulations 
in \citet{2009ApJ...701..945S}.
As a whole, the dependence of $|f_{\rm R0}|$ 
is found to be quite similar to that of 
$\Omega_{\rm m0}$ and $\sigma_{8}$
because $f(R)$ model predicts a higher $\sigma_{8}$
for a fixed amplitude of initial curvature perturbations
and cosmic matter density through a more rapid growth of structure.

\begin{table*}
\caption{
    Expected constraint on $|f_{R0}|$ from Fisher analysis for various statistics. 
    We show the $1$-$\sigma$ error level in units of $10^{-6}$, assuming the fiducial survey parameters;
    $\theta_{G} = 1\, {\rm arcmin}$, $n_{\rm gal}=10\, {\rm arcmin}^{-2}$ and $1,500$ squared degrees.
    We consider both unmarginalized and marginalized cases over 
    $\Omega_{\rm m0}$ and $\sigma_8$. 
    We adopt the maximum multipole for the power and the bispectra
    to be $\ell_\mathrm{max}=2,000$ in the top two rows, while
    the bottom two rows include information on smaller scales up to
    $\ell_\mathrm{max}=8,000$, which probes roughly consistent
    length scales to the peak counts and the MFs.
    \label{tb:fisher}
    }
\begin{tabular}{@{}ccccccccccc}
\hline\hline
 & $P_{\kappa}$ & $B_{\kappa}$ & Peak & MFs & $P_{\kappa}+B_{\kappa}$
 & $P_{\kappa}+$Peak & $P_{\kappa}+$MFs & $P_{\kappa}+B_{\kappa}+$Peak & $P_{\kappa}+B_{\kappa}+$MFs 
 & all
 \\ \hline
 unmarginalized & 0.616 & 1.799 & 0.454 & 0.483 & 0.615 & 0.392 & 0.392 & 0.398 & 0.398 & 0.281 
 \\
 marginalized & 3.069 & 3.980 & 0.565 & 0.936 & 1.400 & 0.546 & 0.802 & 0.533 & 0.753 & 0.409  
 \\ \hline
 unmarginalized & 0.394 & 1.21 & 0.454 & 0.483 & 0.405 & 0.317 & 0.310 & 0.329 & 0.327 & 0.262 
 \\
 marginalized & 1.29 & 2.46 & 0.565 & 0.936 & 0.607 & 0.522 & 0.634 & 0.431 & 0.522 & 0.365
 \\ \hline
\end{tabular}
\end{table*}

The Fisher matrix approach provides a quantitative method to evaluate
the importance of degeneracy among cosmological parameters.
The Fisher matrix $F_{\alpha\beta}$ is given by
\beqa
F_{\alpha \beta} = \sum_{ij} 
{\bf C}_{ij}^{-1} \frac{\partial {\cal O}(x_{i})}{\partial p_{\alpha}}
\frac{\partial {\cal O}(x_{j})}{\partial p_{\beta}},
\label{eq:Fisher}
\eeqa 
where $p_\alpha$ and $p_\beta$ run for cosmological parameters
(i.e., $|f_{\rm R0}|$, $\Omega_{\rm m0}$ and $\sigma_8$
in our case).
Note that we ignore the cosmological dependence of the covariance matrix
in Eq~(\ref{eq:Fisher}).
We here consider two error levels;
un-marginalized error of $|f_{\rm R0}|$ 
and marginalized error of $|f_{\rm R0}|$ 
considering $\Omega_{\rm m0}$ and $\sigma_8$. 
The former corresponds to the case in 
which the parameters $\Omega_{\rm m0}$ and $\sigma_8$ 
(or the amplitude of the primordial fluctuations, $A_s$, 
in realistic situations) are already constrained 
very tightly from independent observations,
while the latter takes into account the effect of parameter degeneracy
on error estimate.
Thus, we can see the importance of parameter degeneracy 
by comparing the un-marginalized and marginalized errors.

Table~\ref{tb:fisher} summarizes 
both un-marginalized and marginalized 
errors of $|f_{\rm R0}|$ for a hypothetical imaging survey
with the sky coverage of 1,500 squared degrees.
The two upper rows correspond to our fiducial case with
$\ell_{\rm max}=2,000$ and $\theta_G=1$ arcmin.
According to the Fisher analysis, 
the power-spectrum analysis results in the most degraded
constraint on $|f_{\rm R0}|$ after marginalization over 
$\Omega_{\rm m0}$ and $\sigma_8$; 
the error level gets $\sim 5$ times larger. 
Although the other cosmic shear statistics do also suffer 
from parameter degeneracy, 
combinations of two or more observables can 
improve the situation quite significantly.
We can confirm in the table that the parameter degeneracy 
is gradually broken by adding statistics one by one.
By properly using 
the four cosmic shear statistics presented in this paper, 
one can provide a better
\textit{marginalized} constraint on $|f_{\rm R0}|$ than 
the \textit{un-marginalized} error expected from 
the power-spectrum analysis alone.
These results would demonstrate the importance of use of different
cosmic shear statistics in upcoming imaging surveys.

So far, our discussion is based on the fiducial analysis.
It is, then, of importance to quantify the constraining power
from different physical scales. Indeed, the maximum multipole 
$\ell_{\rm max}=2,000$ used in the power and the bispectra and 
the smoothing scale $\theta_G=1$ arcmin for the peak counts and 
MFs correspond to somewhat different length scales as we already 
mentioned earlier. We now change the former to $\ell_{\rm max} = 8,000$
to roughly match to $\theta_G=1$ arcmin.
Note that this choice is a bit too aggressive given the larger
uncertainties both in theoretical modelling and measurements.
We show this ideal case only to see the information on the gravity
theory in different statistics from a fair comparison at similar scales.

The resulting constraints on $|f_{\rm R0}|$ are listed in the 
bottom two rows of Table~\ref{tb:fisher}.
The bottom line is the same as the fiducial analysis; the conventional 
power-spectrum analysis exhibits the most notable degradation of the 
constraint on $|f_{\rm R0}|$, and this is mitigated by combining 
more and more statistics. Now, thanks to the small-scale information 
from the power and the bispectra, the error level from each of the
statistics is very similar.

Although Figure~\ref{fig:degen_stat} shows
clear degeneracy among three parameters of 
$|f_{\rm R0}|$, $\Omega_{\rm m0}$ and $\sigma_8$
in cosmic shear statistics, 
the effect of $|f_{\rm R0}|$ 
is not compensated by different $\Omega_{\rm m0}$
and $\sigma_{8}$ exactly.
For instance, the scale-dependent linear growth rate
and the specific feature in the halo mass function in the $f(R)$ model
are quite unique and thus difficult to absorb by a change 
in $\Omega_{\rm m0}$ and $\sigma_8$ 
within the $\Lambda$CDM scenario.

In order to demonstrate this situation more quantitatively, 
we consider the effective 
bias on the $\Omega_{\rm m0}-\sigma_{8}$ plane 
assuming the F6 model is the {\it true} cosmological model 
that governs the universe
but $\Lambda$CDM model is wrongly adopted in the data analysis.
We estimate the bias in parameter estimation as 
\citep{2006MNRAS.366..101H}
\beqa
\delta p_{\alpha} = \sum_{\beta} F_{{\rm GR}, \alpha \beta}^{-1}
\left[ {\cal O}(x_{i}; {\rm F6})-{\cal O}(x_{i}; {\rm fid})\right]
{\bf C}^{-1}_{ij} \frac{\partial {\cal O}(x_{j})}{\partial p_{\beta}},
\label{eq:p_bias}
\eeqa
where $p_{\alpha}=(\Omega_{\rm m0}, \sigma_{8})$,
${\cal O}(x_{i}; {\rm fid})$ represents the assumed $\Lambda$CDM model, while ${\cal O}(x_{i}; F6)$ is the true cosmological model, 
corresponding to the F6 model in this case.
The matrix of $F_{{\rm GR}, \alpha \beta}$ in Eq.~(\ref{eq:p_bias}) is 
the Fisher matrix for $\Omega_{\rm m0}$ and $\sigma_{8}$,
and is the sub-matrix of that in Eq~(\ref{eq:Fisher}).
The Fisher matrix $F_{{\rm GR}, \alpha \beta}$ 
provides the confidence region 
around the fiducial $\Lambda$CDM parameter.
If the difference of the cosmic shear statistics 
between F6 and $\Lambda$CDM models can be explained 
by simply the difference in $\sigma_{8}$,
the bias by Eq~(\ref{eq:p_bias}) would be equal to 
the difference of two values of $\sigma_{8}$ in these models.
More specifically, the bias on $\Omega_{\rm m0}-\sigma_{8}$ plane 
would be equal to 
$(\delta \Omega_{\rm m0}, \delta \sigma_{8})=(0, 
\sigma_{8}({\rm F6})-\sigma_{8}(\Lambda{\rm CDM}))$,
where $\sigma_{8}(\rm F6)$ represents the resulting $\sigma_{8}$
in F6 model, and so on.

\begin{figure}
\centering
\includegraphics[width=0.95\columnwidth,bb=0 0 524 360]
{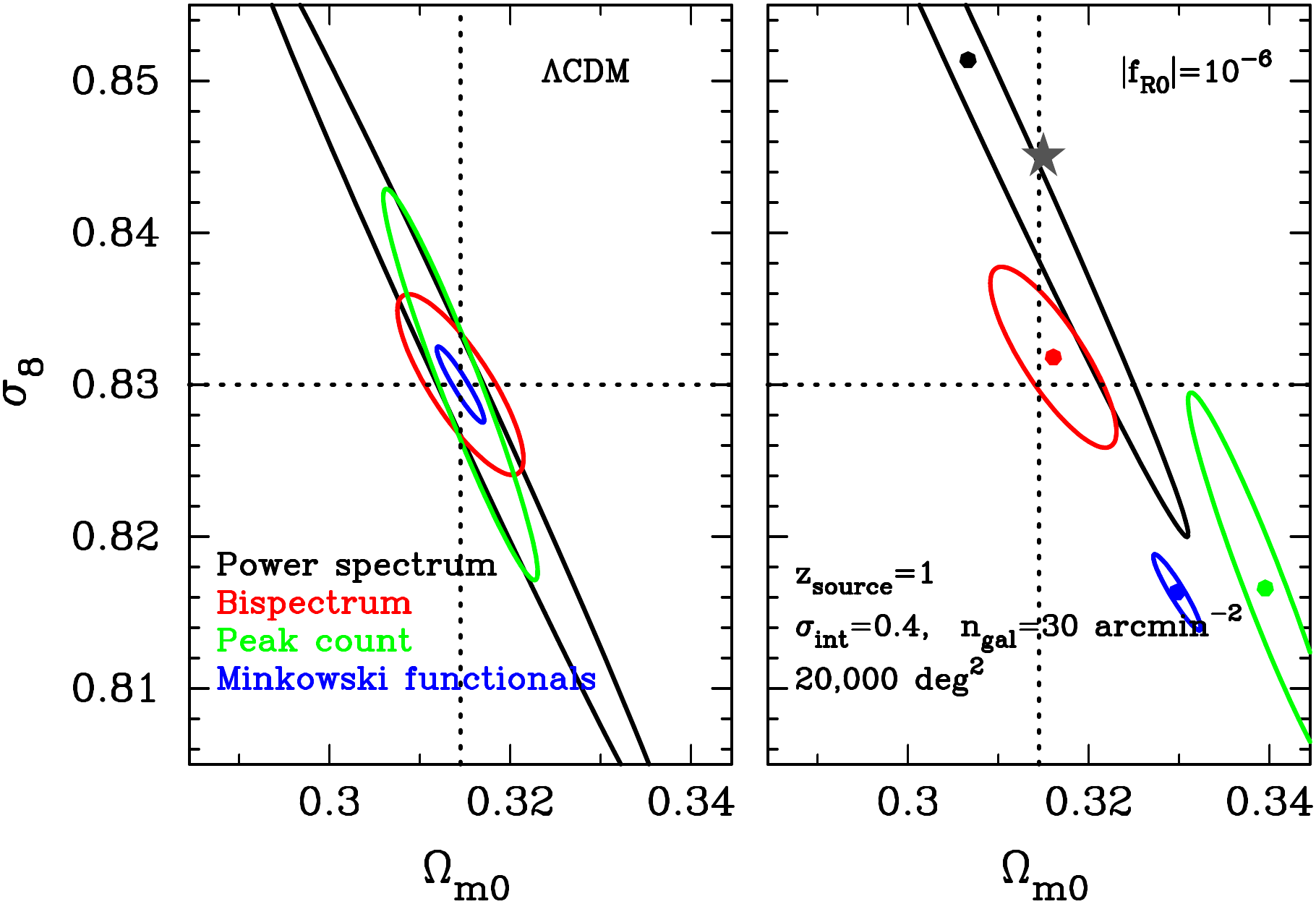}
\caption{
	Consistency among cosmic shear statistics.
	In the left-hand panel, we show the 95\% confidence region
	of $(\Omega_{\rm m0}, \sigma_{8})$
	for $\Lambda$CDM.
	In the right-hand panel, we represent the bias on the 
	$\Omega_{\rm m0}-\sigma_{8}$ plane
	when F6 model is the true model but we use $\Lambda$CDM
	model in parameter constraints.
	The grey star point in the right panel is 
	the expected bias assuming that the difference of statistics between 
	F6 and $\Lambda$CDM can be explained by 
	difference of $\sigma_{8}$.
	In both panels, colored symbols correspond to different
	cosmic shear statistics 
	(black for power spectrum, red for bispectrum, 
	green for peak counts, and blue for MFs).
	In this figure, we assume 
	the sky coverage of 20,000 squared degrees
	and the source number density of 30 ${\rm arcmin}^{-2}$.
	}
\label{fig:consistency_stat}
\end{figure}

Figure~\ref{fig:consistency_stat} shows the result of this analysis.
While we show the standard Fisher analysis within the $\Lambda$CDM
framework in the left-hand panel, we show in the right-hand panel the 
biased estimation in $\Omega_{\rm m0}-\sigma_{8}$ plane induced 
by the difference between F6 and $\Lambda$CDM model.
We show the 95\% confidence region estimated 
from the Fisher matrix with an ideal future survey with 
the sky coverage of 20,000 squared degrees.
The centers of ellipses in the right-hand panel are off from the true
position depicted by the crossing point of the horizontal and the
vertical dotted lines, and the displacement from that point shows 
the effective bias given by Eq.~(\ref{eq:p_bias}).
For simplicity, we employ the size the ellipses the same 
as in the left-hand panel.
We also show by the grey star symbol the expected central 
value if the difference between F6 and 
$\Lambda$CDM can completely be explained by the change in $\sigma_{8}$.

According to the right-hand panel, we find 
that the effect of $f(R)$ gravity on the power spectrum would be
mainly determined by the change of $\sigma_{8}$,
but the other statistics suggest that the difference also propagates to the estimated $\Omega_{\rm m0}$.
Since the convergence bispectrum is less sensitive 
to $|f_{\rm R0}|$, the bias from using it alone 
on $\Omega_{\rm m0}-\sigma_{8}$ plane
would be smaller than the statistical uncertainty in a survey of 
20,000 squared degrees.
This implies that it is difficult to distinguish the F6 model from
$\Lambda$CDM model with bispectrum alone.
Interestingly, peak counts and lensing MFs would predict
higher $\Omega_{\rm m0}$ and lower $\sigma_{8}$ 
if F6 is the true model.
The amount of bias in lensing MFs is smaller than the one in peak counts,
because of different sensitivity of $|f_{\rm R0}|$ 
as shown in Section~\ref{subsec:detectability}.

The result indicates that one can eventually find a clue beyond
the $\Lambda$CDM model by detecting discrepancies in the allowed 
parameter regions from multiple statistics. Notably, a realistic 
analysis of the power and the bispectra up to $\ell_{\rm max}=2,000$
can find this with a high significance for a value of $|f_{\rm R0}|$ 
as small as $10^{-6}$. The additional statistics such as the peak counts
and the MFs would provide an even more promising path towards 
the law of gravity on cosmological scales.

\section{CONCLUSION AND DISCUSSION}
\label{sec:con}

In this paper, we studied the effects of $f(R)$ gravity 
on statistical properties of the weak gravitational lensing field.
For this purpose, we have performed 
$N$-body simulations to investigate structure formation in a universe
under the $f(R)$ model proposed in HS model. 
We then employed ray-tracing method to realize a realistic situation 
of weak lensing measurements in galaxy imaging survey.
In ray-tracing simulations, we have properly taken into account 
the deflection of light along the line of sight and galaxy shape noise.
The large set of these mock lensing catalogs enables us to 
study the information content about $f(R)$ gravity 
in cosmic shear statistics
which have already been conducted in previous imaging surveys.
Our main findings are summarized as follows:

\begin{enumerate}

\vspace{2mm}
\item 
The convergence power spectrum contain information about $f(R)$ gravity because of the scale-dependent linear growth rate 
and the environment-dependence of non-linear gravitational growth.
Assuming the source redshift is set to be 1,
$f(R)$ gravity would enhance the amplitude of the spectrum
at $\ell = 1000$ with a level of $\sim12\%$ and $2\%$ for 
$|f_{\rm R0}|=10^{-5}$ and $10^{-6}$, respectively.
Although the change of convergence power spectrum is expected
given by the difference of $\sigma_{8}$ between $f(R)$ 
and $\Lambda$CDM models, correct understanding of 
the non-linear gravitational growth in $f(R)$ gravity
would be required to determine the amplitude accurately
\citep[e.g., ][]{2014ApJS..211...23Z}.

\vspace{2mm}
\item 
The convergence bispectrum is the lowest-order non-Gaussian 
information in weak lensing field.
We find that it can change by 
$\sim10\%$ with the model of $|f_{\rm R0}|=10^{-5}$
and the dependence on the triangle configuration in Fourier space is weak.
However, the information of $f(R)$ gravity in convergence bispectrum
would be less important than that in the power spectrum, because 
the change of amplitude would be smaller than the statistical uncertainty
even in an upcoming survey with a sky coverage of 20,000 squared degrees.
Our results indicate that the information from the convergence 
bispectrum should 
be used to break the degeneracy between 
$|f_{\rm R0}|$ and the present amplitude of matter fluctuations 
$\sigma_{8}$ in the convergence power spectrum.
This is consistent with the previous investigation in 
\citet{2011JCAP...11..019G}.

\vspace{2mm}
\item 
Peak counts in a reconstructed smooth convergence field 
are expected to be informative for constraining the nature of gravity, 
because the modification of gravity can affect 
the abundance of massive dark matter halos.
We find that the number density of peaks can be affected 
by the presence of extra scalar degree of freedom with $|f_{\rm R0}|=10^{-5}$ with a level of a few percents.
Besides the peaks with high height caused by isolated massive objects 
along a line of sight, the peaks with S/N of $\sim1$
can be useful to distinguish the $f(R)$ model with  $|f_{\rm R0}|=10^{-5}$ from GR. 
This information at intermediate peak height is similar to the effect of 
changing $\sigma_{8}$ in $\Lambda$CDM (c.f.~Figure~\ref{fig:degen_stat}; lower left-hand panel),
but again it is difficult to exactly compensate the difference
of peak counts in $f(R)$ gravity and GR by
varying $\sigma_{8}$. 
 
\vspace{2mm}
\item
MFs are an interesting statistic 
to extract non-Gaussian information from 
a given random field.
Previous study \citep{2015PhRvD..92f4024L} has investigated 
the possibility of using them to constrain on $f(R)$ model, while 
we improve the analysis by 
considering realistic observational situations.
We find that lensing MFs in a reconstructed smooth convergence field
show $2-3\%$ differences between two cases of 
$|f_{\rm R0}|=10^{-5}$ and 0 (or GR).
The effect of $|f_{\rm R0}|$ in lensing MFs are reduced 
by the presence of shape noise and projection effect compared to 
the previous work, showing our approach with realistic ray-tracing simulations would be essential to predict them.
Although $f(R)$ model with $|f_{\rm R0}|=10^{-5}$ 
would affect the lensing MFs with only a few percent,
MFs are still useful to constrain $f(R)$ gravity because of 
their small statistical uncertainty.
However, the constraining power of $|f_{\rm R0}|$ in lensing MFs
would be strongly dependent on the smoothing scale in reconstruction.

\vspace{2mm}
\item 
Among the four statistics, 
the convergence power spectrum, peak counts and lensing MFs have 
a similar sensitivity to $f(R)$ gravity 
in typical ground-based imaging surveys
if the small-scale clustering of lensing fields 
at $\simgt1$ arcmin can be properly modelled. 
When we apply a larger smoothing to 
match the probed scale effectively to $\ell\simlt2,000$,
the non-Gaussian statistics have shown similar sensitivity 
to $f(R)$ gravity.
Nevertheless, 
the information of peak counts and lensing MFs can be improved 
by a factor of $2-3$ when one can reduce the shape noise contaminations
in a smoothed convergence map by increasing the source number density.
In terms of degeneracy among cosmological parameters, 
the convergence power spectrum has the  
strongest degeneracy between $|f_{\rm R0}|$ and $\sigma_{8}$,
while peak counts and lensing MFs would show a different degeneracy.
Therefore, a complete and accurate understanding of peak counts and lensing MFs would be 
helpful to break the degeneracy between modified gravity 
and the concordance $\Lambda$CDM parameters.
Note that the convergence bispectrum can be an 
unbiased indicator of $\Omega_{\rm m0}$ and $\sigma_{8}$
because of weak dependence of $|f_{\rm R0}|$.

\end{enumerate}

Our findings are important for
constraining the nature of gravity with weak lensing measurement.
There still remain, however, crucial issues 
on the cosmic shear statistics proposed in this paper.

Although the peak counts and lensing MFs 
can be used to extract cosmological information
beyond the two-point statistics, at 
the crucial length scales of structure probed by them,
perturbative approaches break down because of 
the non-linear gravitational growth 
\citep{2002ApJ...571..638T, Petri2013}.
In order to sample a Likelihood function for a wide range of cosmological parameters, 
we require accurate theoretical predictions of 
the non-local statistics in convergence map beyond perturbation methods
\citep[e.g.,][]{Matsubara2003}.
A simplest approach to build the predictions for various cosmological models
is to use a large set of cosmological $N$-body simulations 
\citep{2014ApJ...786...43S, 
2015PhRvD..91f3507L, 2015PhRvD..91j3511P}, 
while there exists a more flexible approach to predict the non-local statistics \citep[e.g.,][]{2015A&A...576A..24L}.
Another important issue is theoretical uncertainties 
associated with baryonic effects.
Previous studies
\citep[e.g.,][]{2013MNRAS.434..148S,2013PhRvD..87d3509Z}
explored the effect of including baryonic components 
to the two-point correlation of cosmic shear 
and consequently to cosmological parameter estimation.
\citet{2015ApJ...806..186O} have studied baryonic effects
on peak counts and lensing MFs with hydrodynamic simulations
under GR.
Although the baryonic physics would play an important role in 
the regions where GR should be recovered, 
the modification of gravity might affect the large-scale structure.
Since the peak count and lensing MFs would involve in 
the cosmological information of various structures 
in the Universe in complex way, 
we need to develop accurate modelling of cosmic shear statistics
incorporated with both modifications of gravity and baryonic effects.
The statistical properties and the correlation 
of source galaxies and lensing structures
are still uncertain but could be critical when making lensing mass maps.
For example, source-lens clustering 
\citep[e.g.,][]{2002MNRAS.330..365H}
and the intrinsic alignment \citep[e.g.,][]{2004PhRvD..70f3526H}
are likely to affect the information content of 
$f(R)$ gravity in cosmic shear statistics. 
There is a possibility that these two effects can induce
the systematic effect on reconstruction of lensing mass maps
\citep[e.g.,][]{2016MNRAS.463.3653K}.
We plan to address these important issues in future works.

Upcoming imaging surveys would provide imaging data of billions of galaxies at $z\sim1$.
Detailed statistical analyses of these precious data sets 
can reveal matter density distribution in the Universe regardless 
of which cosmic matter is luminous and dark.
Since matter contents in the Universe can be evolved by 
nonlinear gravitational growth, a map of matter distribution observed 
in future would be a key to understand the nature of gravity.
Cosmological tests of GR with imaging surveys
are just getting started and the present work in this paper 
would a useful step in understanding the nature of gravity 
with future weak lensing data.

\section*{acknowledgments}
MS is supported by Research Fellowships of the Japan Society 
for the Promotion of Science (JSPS) for Young Scientists.
BL is supported by STFC Consolidated Grant No. ST/L00075X/1 and 
No. RF040335.
YH is supported by Academia Sinica, Taiwan. 
Numerical computations presented in this paper were in part carried out
on the general-purpose PC farm at Center for Computational Astrophysics,
CfCA, of National Astronomical Observatory of Japan.
Data analyses were [in part] carried out on common use data analysis computer system at the Astronomy Data Center, 
ADC, of the National Astronomical Observatory of Japan.

\bibliographystyle{mnras}
\bibliography{bibtex}

\appendix
\section{The detailed configuration of triangles in lensing bispectrum}

We provide the information of lensing bispectrum used in Figure~\ref{fig:degen_stat}.

\begin{table}
\centering
\caption{
	The configuration of triangles in lensing bispectrum shown in Figure~\ref{fig:degen_stat}.
	The ordering of ID corresponds to the data in Figure~\ref{fig:degen_stat} from left to right.  
	}
\begin{tabular}{@{}lccccl}
\hline
\hline
ID & $\ell_1$ & $\ell_2$ & $\ell_3$ &\\ \hline 
Equilateral & & &\\ \hline
1 & 119.4 & 119.4 & 119.4 &\\
2 & 242.4 & 242.4 & 242.4 &\\ 
3 & 345.5 & 345.5 & 345.5 &\\ 
4 & 492.4 & 492.4 & 492.4 &\\ 
5 & 701.7 & 701.7 & 701.7 &\\ 
6 & 1000.0 & 1000.0 & 1000.0 &\\ 
7 & 1425.1 & 1425.1 & 1425.1 &\\ 
\hline
Isosceles & & &\\ \hline
1 & 119.4 & 119.4 & 170.1 &\\ 
2 & 119.4 & 119.4 & 242.4 &\\ 
3 & 170.1 & 170.1 & 242.4 &\\
4 & 170.1 & 170.1 & 345.5 &\\ 
5 & 242.4 & 242.4 & 345.5 &\\ 
6 & 242.4 & 242.4 & 492.4 &\\ 
7 & 345.5 & 345.5 & 492.4 &\\ 
8 & 345.5 & 345.5 & 701.7 &\\ 
9 & 492.4 & 492.4 & 701.7 &\\ 
10 & 492.4 & 492.4 & 1000.0 &\\ 
11 & 701.7 & 701.7 & 1000.0 &\\ 
12 & 701.7 & 701.7 & 1425.1 &\\ 
13 & 1000.0 & 1000.0 & 1425.1 &\\
\hline
\end{tabular}
\end{table}

\begin{table}
\centering
\caption{Similar to Table~A1, but for scalene triangles.}
\begin{tabular}{@{}lccccl}
\hline
\hline
ID & $\ell_1$ & $\ell_2$ & $\ell_3$ &\\ \hline
Scalene & & &\\ \hline
1 & 119.4 & 170.1 & 242.4 &\\ 
2 & 119.4 & 170.1 & 345.5 &\\
3 & 119.4 & 242.4 & 345.5 &\\ 
4 & 119.4 & 242.4 & 492.4 &\\ 
5 & 119.4 & 345.5 & 492.4 &\\ 
6 & 119.4 & 492.4 & 701.7 &\\ 
7 & 119.4 & 701.7 & 1000.0 &\\ 
8 & 119.4 & 1000.0 & 1425.1 &\\ 
9 & 170.1 & 242.4 & 345.5 &\\ 
10 & 170.1 & 242.4 & 492.4 &\\ 
11 & 170.1 & 345.5 & 492.4 &\\ 
12 & 170.1 & 345.5 & 701.7 &\\ 
13 & 170.1 & 492.4 & 701.7 &\\ 
14 & 170.1 & 701.7 & 1000.0 &\\ 
15 & 170.1 & 1000.0 & 1425.1 &\\ 
16 & 242.4 & 345.5 & 492.4 &\\ 
17 & 242.4 & 345.5 & 701.7 &\\ 
18 & 242.4 & 492.4 & 701.7 &\\ 
19 & 242.4 & 492.4 & 1000.0 &\\ 
20 & 242.4 & 701.7 & 1000.0 &\\ 
21 & 242.4 & 1000.0 & 1425.1 &\\ 
22 & 345.5 & 492.4 & 701.7 &\\ 
23 & 345.5 & 492.4 & 1000.0 &\\ 
24 & 345.5 & 701.7 & 1000.0 &\\ 
25 & 345.5 & 701.7 & 1425.1 &\\ 
26 & 345.5 & 1000.0 & 1425.1 &\\ 
27 & 492.4 & 701.7 & 1000.0 &\\ 
28 & 492.4 & 701.7 & 1425.1 &\\ 
29 & 492.4 & 1000.0 & 1425.1 &\\ 
30 & 701.7 & 1000.0 & 1425.1 &\\ 
\hline
\end{tabular}
\end{table}

\end{document}